\documentclass[twocolumn]{aastex631}

\hypersetup{linkcolor=blue,citecolor=blue,filecolor=blue,urlcolor=blue}

\usepackage{lmodern}
\usepackage{amsmath}
\usepackage{amssymb}
\usepackage{bm}
\usepackage{hyperref}
\usepackage{url}
\usepackage{xcolor}
\usepackage[demo]{rotating}
\usepackage{multirow}

\DeclareMathOperator{\EX}{\mathbb{E}}

\newcommand{\x}{\mathbf{x}}

\usepackage{orcidlink}
\newcommand{\angstrom}{\text{\normalfont\AA}}
\accepted{October 25, 2024}

\shortauthors{Doorenbos, Sextl et al.}
\shorttitle{Galaxy Properties from photometric images}

\begin{document}

\title{Galaxy spectroscopy without spectra:\\
Galaxy properties from photometric images with conditional diffusion models}

\correspondingauthor{Lars Doorenbos, Eva Sextl}
\email{lars.doorenbos@unibe.ch, sextl@usm.lmu.de}

\author{Lars Doorenbos\textsuperscript{*} \orcidlink{0000-0002-0231-9950}}
\affiliation{AIMI, ARTORG Center, University of Bern, Murtenstr. 50, CH-3008 Bern, Switzerland}

\author{Eva Sextl\textsuperscript{*} \orcidlink{0009-0001-5618-4326}}
\affiliation{Universit\"ats-Sternwarte, Fakult\"at f\"ur Physik, Ludwig-Maximilians Universit\"at M\"unchen, Scheinerstr. 1, 81679 M\"unchen, Germany}

\author{Kevin Heng \orcidlink{0000-0003-1907-5910}}
\affiliation{Universit\"ats-Sternwarte, Fakult\"at f\"ur Physik, Ludwig-Maximilians Universit\"at M\"unchen, Scheinerstr. 1, 81679 M\"unchen, Germany}

\author{Stefano Cavuoti \orcidlink{0000-0002-3787-4196}}
\affiliation{INAF - Astronomical Observatory of Capodimonte, Salita Moiariello 16, I-80131 Napoli, Italy}
\affiliation{INFN - Sezione di Napoli, via Cinthia 9, 80126 Napoli, Italy}

\author{Massimo Brescia \orcidlink{0000-0001-9506-5680}}
\affiliation{Department of Physics, University Federico II, Strada Vicinale Cupa Cintia, 21, 80126 Napoli, Italy}
\affiliation{INAF - Astronomical Observatory of Capodimonte, Salita Moiariello 16, I-80131 Napoli, Italy}
\affiliation{INFN - Sezione di Napoli, via Cinthia 9, 80126 Napoli, Italy}

\author{Olena Torbaniuk \orcidlink{0000-0003-4465-2564}}
\affiliation{Department of Physics and Astronomy `Augusto Righi', University of Bologna, via Piero Gobetti 93/2, 40129 Bologna, Italy}

\author{Giuseppe Longo \orcidlink{0000-0002-9182-8414}}
\affiliation{Department of Physics, University Federico II, Strada Vicinale Cupa Cintia, 21, 80126 Napoli, Italy}

\author{Raphael Sznitman \orcidlink{0000-0001-6791-4753}}
\affiliation{AIMI, ARTORG Center, University of Bern, Murtenstr. 50, CH-3008 Bern, Switzerland}

\author{Pablo Márquez-Neila \orcidlink{0000-0001-5722-7618}}
\affiliation{AIMI, ARTORG Center, University of Bern, Murtenstr. 50, CH-3008 Bern, Switzerland}

\received{June 24, 2024}

\begin{abstract}
Modern spectroscopic surveys can only target a small fraction of the vast amount of photometrically cataloged sources in wide-field surveys. Here, we report the development of a generative AI method capable of predicting optical galaxy spectra from photometric broad-band images alone. This method draws from the latest advances in diffusion models in combination with contrastive networks. We pass multi-band galaxy images into the architecture to obtain optical spectra. From these, robust values for galaxy properties can be derived with any methods in the spectroscopic toolbox, such as standard population synthesis techniques and Lick indices. When trained and tested on 64 × 64-pixel images from the Sloan Digital Sky Survey, the global bimodality of star-forming and quiescent galaxies in photometric space is recovered, as well as a mass-metallicity relation of star-forming galaxies. The comparison between the observed and the artificially created spectra shows good agreement in overall metallicity, age, Dn4000, stellar velocity dispersion, and E(B-V) values. Photometric redshift estimates of our generative algorithm can compete with other current, specialized deep-learning techniques. Moreover, this work is the first attempt in the literature to infer velocity dispersion from photometric images. Additionally, we can predict the presence of an active galactic nucleus up to an accuracy of $82\,\%$. With our method, scientifically interesting galaxy properties, normally requiring spectroscopic inputs, can be obtained in future data sets from large-scale photometric surveys alone. The spectra prediction via AI can further assist in creating realistic mock catalogs.
\end{abstract}

\footnotetext[0]{\textsuperscript{*}The first two authors share the first authorship and are in alphabetical order.}



\section{Introduction} \label{sec:intro}

Astrophysics, like many other sciences, is currently undergoing a significant transformation due to the avalanche of high-quality data from various sky surveys \citep{Borne2010, Bell2009}. Merely forty years ago, astronomical image data sets were measured in kilo- or megabytes. Twenty years ago, by the beginning of the 21st century, data releases for the first large-scale survey, the Sloan Digital Sky Survey (SDSS), started. Other well-known surveys such as Pan-STARRS \citep{Kaiser2002}, DESI \citep{DESIcoll1, DESIcoll2}, and Euclid \citep{Racca2016} followed. DESI alone now has captured more galaxies than 10 years of SDSS \citep{Setton2023}. The currently built Vera C. Rubin Observatory in Chile is designed to collect 20 terabytes per night over the time span of 10 years \citep{VeraRubinObs2019}. Yet, most objects are only captured via photometry or photometric imaging. Spectra are only available for a small portion of the captured galaxies due to the extended duration needed for exposure and the limited capacity of spectroscopic instruments to handle multiple measurements simultaneously \citep{Kremer2017}. Furthermore, the magnitude limits for spectroscopy are much brighter, thus preventing spectroscopic observations of faint galaxies. For instance, Legacy Survey of Space and Time (LSST) investigations will likely acquire spectra for less than 1\% of the galaxies involved \citep{Matheson2013}. 

Greatly simplified, the overall goal of photometry is to map observed colors to galaxy properties. When only apparent magnitudes in several broad-band filters are available for a galaxy, performing a panchromatic SED fitting (from UV to IR) has become a popular approach. The best suitable spectral energy distribution is chosen from a large sample of pre-computed templates, which assume, among others, different star formation and metal enrichment histories \citep{Mitchell2013, Yuan2020}. The use of state-of-the-art publicly available codes is discussed in \citet{Lower2020, Mierlo2023}. The SED fitting method allows for the evaluation of critical characteristics like the star formation rate (SFR) and stellar mass of a galaxy, which are crucial for comprehending their formation and evolution. Normally, these methods use solely color information, not images with morphological features (i.e., the brightness distribution across the galaxy). With such reconstructed SEDs, it is not possible to make trustworthy and accurate age or metallicity claims, as was recently shown in \citet{Nersesian2024}. Even the use of additional narrow-band filters to given broad-bands does not seem to improve the recovery of galaxy parameters beyond stellar mass and SFR \citep{Csizi2024}. Determining stellar age, stellar metallicity, and dust properties with photometry alone seem rather hopeless.

With the increasing integration of Artificial Intelligence (AI) in astronomy, new avenues have opened up for deriving parameters from photometric data with machine learning techniques. \citet{Wu2023} predicted stellar atmospheric parameters like effective temperature from stellar photometric images, and \citet{Chen2021,doorenbos2022ulisse} succeeded in recognizing active galactic nuclei (AGN) with a deep neural network that takes photometric magnitudes as input. Several classification tasks of celestial objects, which would traditionally require spectroscopic data, will soon be fully automated using only photometric bands \citep{Zeraatgari2024}. In this paper, we combine our proposed generative AI (GenAI) system with large quantities of photometric images to go a step further toward an all-embracing use of the upcoming full-sky surveys.

We present a pilot study of predicting optical spectra directly from photometric broad-band images in the SDSS survey. We show that their spectral resolution and overall quality are sufficient to analyze them with common spectroscopic tools. In doing so, we can recover interesting physical parameters such as the population age or mean metallicity. This could be a classical absorption feature analysis (i.e. Lick indices) or a full-spectral fitting code as part of stellar population synthesis. 

Unlike previous attempts in the literature, we choose to make a detour over the generation of optical spectra and \emph{not} predict the physical quantities directly from the images. Therefore, we can detach ourselves from answering unsolved questions such as which full-spectral fitting code performs best \citep{Woo2024}. Once the spectrum is created from the generative AI, the choice of the spectroscopic analysis method is left to the user. It must only be ensured that the quality and the information content of the predicted spectra are similar to the real, observed ones. We believe that this freedom in subsequent research questions is worth the higher computational cost for a generative AI. In section \ref{sec:data}, we describe the utilized SDSS data and introduce our machine learning pipeline and implementation in section \ref{sec:ml}. We use a diverse tool set to evaluate the predicted spectra, for which we explain the details in section \ref{sec:evaluation}. Major results are presented in section \ref{sec:results}. A final summary with an outlook is found in \ref{sec:summ}. Further technical comments on the algorithms are given in the appendix \ref{sec:appendix}. Additional thoughts on the test set are also found there.

\section{Data} \label{sec:data}

The significance of data in machine learning cannot be overstated. The type and quantity of data provided to an algorithm play a crucial role in its ability to extract information and generate accurate results. Creating artificial spectra in a spectral resolution suitable for follow-up analysis likely requires a large dataset with hundreds of thousands of entries. The natural first choice for this is the Sloan Digital Sky Survey. In its third phase in 2014, it encompassed more than one-third of the entire celestial sphere and is freely accessible\footnote{\url{https://live-sdss4org-dr12.pantheonsite.io/scope/}}. Numerous research groups have already applied a variety of machine learning techniques to SDSS in order to answer scientific questions about galaxies, quasars, and various other celestial objects (\citealt{Hoyle2015, Si2021, Curti2022, Miller2015, brescia2015automated} to name a few). We aim to build further upon this work. 

\subsection{Multiband images}

We utilize pre-processed broadband images from the dataset made available\footnote{ \url{https://deepdip.iap.fr/\#item/60ef1e05be2b8ebb048d951d}} by \citet{Pasquet2019} for their work on photometric redshift estimation with a deep convolutional neural network. For each broadband in SDSS, the galaxy’s brightness is captured in an image. Their total dataset contains $659\,857$ galaxies of the 12th Data Release (DR12) of the Sloan Digital Sky Survey \citep{SDSSDR12} in a redshift range of $0 < z < 0.7$. For each galaxy, astrometrically calibrated imaging data in u, g, r, i, and z filters with $0.396^ {\prime \prime }$/pixel sampling is avaiflable through the automated data-pipeline of SDSS \citep{Padmanabhan2008, Stoughton2002}. Further re-sampling and stacking of the obtainable broad-band images by \citet{Pasquet2019} leads to a final 64×64×5 data cube centered on each spectroscopic target. For our purposes, a further cut in redshift is made later on. 

\begin{figure}[htb]
        \includegraphics[width=\columnwidth]{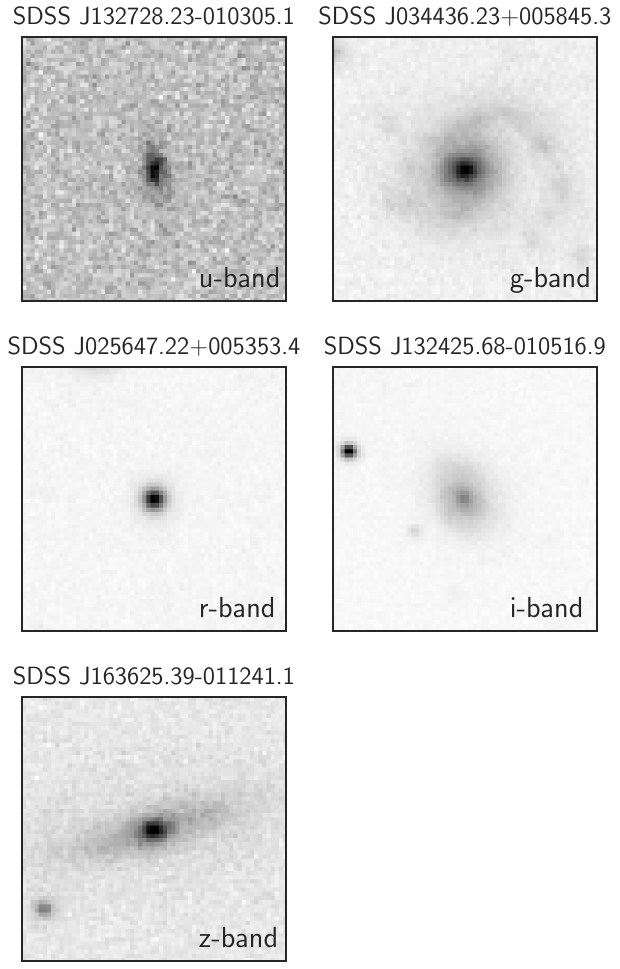}
    \caption{Visualization of photometric images of \emph{different} galaxies in the training set, which serves as the input for the machine learning algorithm. The corresponding galaxy name can be found at the top of the image and the depicted band is written on each image. We chose to visualize diverse objects to better emphasize the variety of galaxy images in the training set.} \label{fig:example_images}
\end{figure}

Example images can be found in figure \ref{fig:example_images}. It should be emphasized that images with adjacent objects (in either the forefront or background) are not removed from the sample. As a result, together with the size, morphology, and surface brightness, information about the environment is also fed into the generative model \citep{Dey2022}. Thus, more information is available to break various degeneracies.

\subsection{Galaxy spectra}

After obtaining the images, the corresponding optical galaxy spectra and their labels were extracted from the SDSS database. For such large queries, the most efficient way is using CasJobs \citep{CasServer2008}, a flexible, advanced SQL-based interface. We obtain galaxy spectra in the spectroscopic redshift range of $0.05 < z \leq 0.15$, which show no problematic flags\footnote{\url{https://live-sdss4org-dr16.pantheonsite.io/tutorials/flags}} and reasonable Petrosian radii and de Vaucouleurs/exponential surface brightness fits. The complete query code is shown in Appendix~\ref{app:sql}. This final sample contains $270\,621$ image-spectra pairs spanning a wide range of morphology classes, such as spiral galaxies with bulge and/or bar components, ellipticals, and irregular galaxies at low redshift with their distinct spectral features. To minimize errors caused by aperture effects, more than 20\% of the total light emitted by each galaxy should be captured by the fibres. This fraction is guaranteed by implementing the lower redshift cut of 0.05 in our sample selection \citep{Kewley2005}. Yet, the derivation of SFRs would still be heavily affected by the aperture in this redshift bracket \citep{Duarte2017}, so we refrain from obtaining them.

A further word of caution needs to be said in this regard. The aperture effect inherently found in the SDSS data itself can technically also lead to false classifications. For instance, a galaxy might be erroneously categorized as retired based on its central spectrum, when in reality, it harbors active star formation in its outer regions not captured by the SDSS fiber. This discrepancy highlights a fundamental mismatch between the photometric and spectroscopic data in SDSS which is hard to come by: while photometric images capture the entire visible extent of galaxies, the spectra used for training are limited to the fiber-covered regions. While the bias is expected to be relatively consistent across the sample within our chosen redshift regime, the model probably cannot be reliably applied to predict spectra of galaxies outside this range. It is beyond the scope of this paper to mitigate this effect.

\begin{sidewaysfigure*}
    \raggedleft 
    \vspace*{10cm}
        \includegraphics[width=\columnwidth]{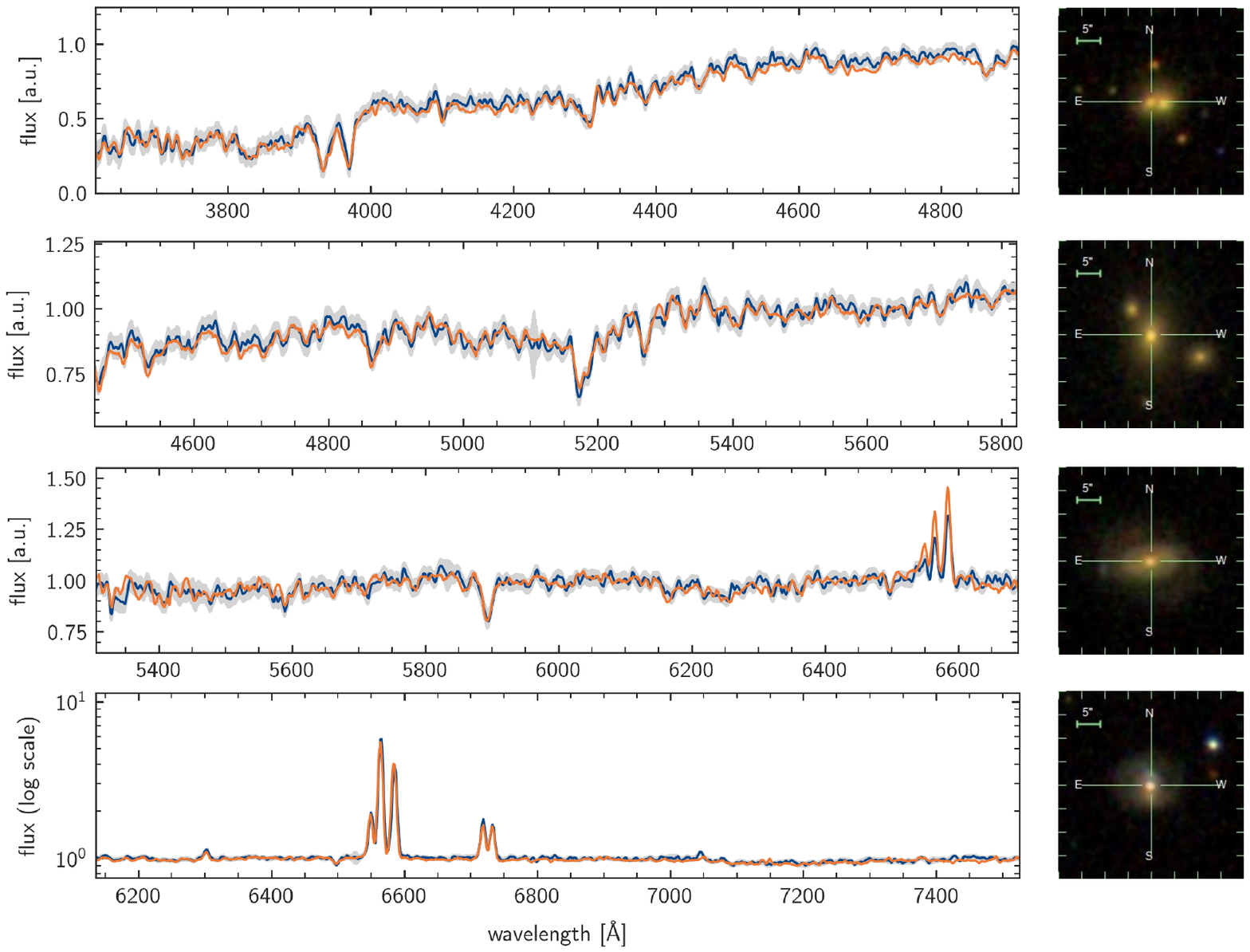}
    \caption{An overview of different predicted spectra (orange) and their associated smoothed observed spectra (blue) in the rest-frame from the test set. Grey-shaded areas depict the corresponding $1-\sigma$ error bars of the smoothed observed spectra. The galaxies show different morphologies (according to the Galaxy Zoo classification: uncertain, elliptical, uncertain, spiral) as visible in the corresponding SDSS tri-color images from the inner three bands on the right. These jpeg images are available on the Skyserver and only serve as an illustration for the reader; the algorithm uses 5-band images. The first example is a possible merger; the third example is classified as LINER by \citet{Brinchmann2004}, and the galaxy at the bottom is best described as star bursting. The galaxy names read (from top to bottom): SDSS J205600.32-053137.8, SDSS J150109.86+472039.6, SDSS J105728.22+065954.4, SDSS J025019.52-070223.4 } \label{fig:example_spec}
\end{sidewaysfigure*}

We split the data into $268\,603$ samples for training the algorithm, $512$ for validation/fine-tuning the model's performance, and $1\,506$ as the test set. We set apart these fractions of the data for the validation and test sets due to the large overall size of the dataset and the large number of computationally involved analyses of the generated spectra we performed. Nonetheless, a test set with over $1\,500$ objects allows us to draw robust conclusions in the following sections, and small ratios for testing and validation are standard practice in deep learning when dealing with big datasets and/or computationally involved evaluations~\citep{Amari1997,Ng2017}. Moreover, in absolute terms, our validation and test sizes match those used in deep learning practice, for instance, in semantic segmentation~\citep{cordts2016cityscapes}. The test dataset was never used in training; instead, it is used to assess the model's performance on unseen data. Figure \ref{fig:example_spec} shows some artificial spectra from the test set compared to their observed counterparts. 

The morphological categories ``elliptical'', ``spiral'', and ``uncertain'' for each galaxy in the test set, which are used later on, are determined by the citizen science project Galaxy Zoo \citep{Lintott2008, Lintott2011}. Galaxies were labeled as ``uncertain'' if their images were not clearly voted on as spiral or elliptical. These galaxies are most likely composite bulge-disk systems in which neither the bulge nor disk overshadows the other according to \citet{Schawinski2014}. In this context, it should also be mentioned that a clear identification of merger systems via images is difficult as galaxies can appear to be isolated galaxies in the image (lacking visual features like tidal features) but appear to have undergone a recent merger when further investigated \citep{Nevin2019}. Therefore, analysis of the galaxies with respect to the categories merger/no merger was omitted. We reduced the resolution of the spectra from the original $R \sim 2000$ to $R \sim 1500$ at $5000 \, \angstrom$ due to the high computing capacity needed. Nonetheless, this is still larger than comparable studies \citep{holwerda2021predicting,Wu2020}. These smoothed spectra are used during training; therefore, the predicted spectra also show this reduced spectral resolution. As final preparatory work, the spectra are interpolated and tailored to 1 Angstrom steps in the range of $4000-8499 \, \angstrom$. Each spectrum is then normalized to a value of 1 between $6900-6950 \, \angstrom$ (R-band) rest-frame wavelength. This region does not contain prominent absorption or emission features and is, therefore, suited for the task, as we want the overall continuum to be scaled. Other possible wavelength regions would be $4400-4450 \, \angstrom$ (B-band) or $5500-5550 \, \angstrom$ (V-band). This does not change the results of the upcoming analysis, especially not those of the full-spectral fitting \citep{Sextl2023}.
For the spectral fits in section \ref{sec:eval1}, the spectrum to be fitted and the model templates always receive such a scaling for numerical stability.  

Now-and-then occurring skylines in the training data are not removed, and the spectra are \emph{not} shifted in the rest-frame. We also decided against removing galactic extinction in the images and the spectra since the overall reddening is rather small: $85\%$ of our sample has E(B\nobreakdash-V) values lower than $0.05$ \citep{Pasquet2019}. 

\section{Procedure} \label{sec:ml}

We frame the problem of predicting galaxy spectra from photometry as modeling the conditional distribution of spectra given an image\footnote{Parts of this section are based on \citet{Doorenbos2022}, presented at the NeurIPS 2022 workshop on Machine Learning and the Physical Sciences.}. Specifically, we want to approximate the empirical distribution 
\begin{equation}
    p(\text{spectrum}\mid\text{image}),
\end{equation}
which is defined by the training data, using a neural network. However, directly learning this distribution over high-resolution spectra is both challenging and computationally expensive.

Instead, we follow recent works on high-resolution image synthesis e.g., \citet{blattmann2023align,ho2022cascaded,saharia2022photorealistic} and decompose the problem into two parts. First, we learn the simpler distribution of low-resolution spectra conditioned on images, 
\begin{equation}
    p(\text{low-res spectrum}\mid\text{image}) \equiv p_{lr}.
\end{equation}
Second, we learn the image-conditional distribution over full-resolution spectra, with an additional condition on the corresponding low-resolution spectrum,
\begin{equation}
    p(\text{spectrum}\mid\text{image, low-res spectrum}) \equiv p_{sr}.
\end{equation}

By combining the two, we can generate a spectrum for a given image by drawing a sample from $p_{lr}$ and then using it as the condition for $p_{sr}$. This effectively upsamples the low-resolution spectrum to the original resolution, which is fit for analysis.
In practice, we learn both distributions with a conditional diffusion model (CDM), which is the current state-of-the-art in generative modeling \citep{Ho2020,Rombach2022}. Generative models are a class of data-driven algorithms that model the probability distribution of observed data and generate new samples resembling the training data. These models have found success in various domains by offering the ability to easily generate new data instances that closely mirror the characteristics of the training set~\citep{baranchuk2021label,de2022next,zhang2021datasetgan}.

While CDMs can generate realistic samples that closely mirror the characteristics of the training set, they do not allow for density estimation. Consequently, sampling from the CDMs results in multiple possible spectra for a given object without information on their likelihood. Nonetheless, we need to select one of the spectra for our subsequent analyses. To decide which spectrum we select for follow-up evaluation, we use multimodal contrastive learning \citep{Chen2020} as a heuristic to find high-likelihood samples of the learned distribution, which has proven to work well in practice \citep{ramesh2021zero}. Multimodal contrastive learning is a representation learning algorithm that learns informative features from a dataset without having access to any labels.

Specifically, multimodal contrastive learning learns to map images and spectra into a shared representation space, where images and spectra with similar representations are likely to belong to the same object. We rank the generated spectra based on the similarities between their representations and that of the original image, then select the best-matching samples.

Our full method for predicting spectra from photometry begins by sampling several 563-dimensional spectra for an image with the low-resolution CDM. We select the three best synthetic spectra according to the low-resolution contrastive network. Then, we generate five full-resolution spectra for each of the selected low-resolution spectra. Finally, we select the best-matching spectrum with the full-resolution contrastive network, giving us the final synthetic spectrum for the object. 

A visualization of our pipeline is provided in Figure~\ref{fig:method}. We provide further technical details in Appendix~\ref{app:tech}.

\begin{figure}
    \centering
    \includegraphics[width=0.85\linewidth]{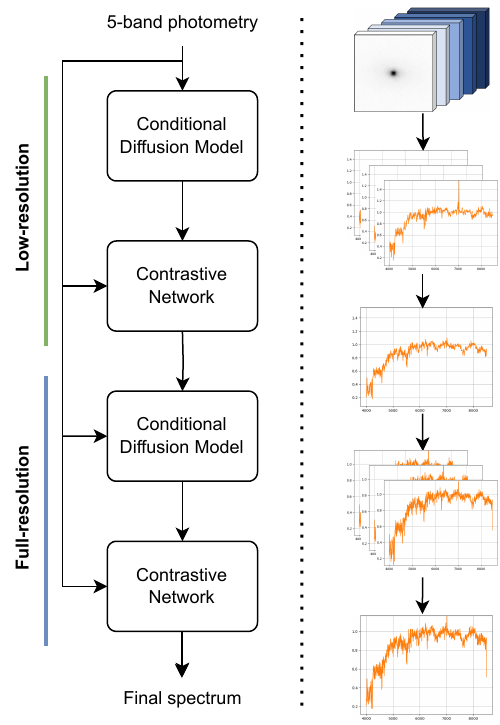}
    \caption{Visualization of our pipeline. We use a conditional diffusion model to generate candidate low-resolution spectra for the given image and select the best matching candidates with a contrastive network. Then, we use a conditional diffusion model to upsample the candidate spectra to full-resolution candidates and select the final predicted spectrum with the second contrastive network.}
    \label{fig:method}
\end{figure}

\subsection{Implementation details}

We train all networks using Adam \citep{kingma2014adam} with a learning rate of $10^{-4}$, with the contrastive networks using a weight decay of $10^{-3}$. We use a batch size of 512 for the low-resolution experiments and 224/340 for the full-resolution CDM and contrastive network, respectively. The CDM uses 250 timesteps with a cosine variance schedule, exponential moving average with $\alpha=0.9999$, a ResNet-18 \citep{he2016deep} for the image encoder $\tau_\theta$ and a 1D U-net \citep{ronneberger2015u} for the denoising autoencoder. The contrastive networks use a ResNet-50 \citep{he2016deep} for the image encoder and a 1D ResNet encoder for the spectra, both with a latent dimensionality of 128. 

We standardize the images by channel and predict the logarithm of the spectra so that the ranges of values better match the Gaussian noise used by the CDM. We apply data augmentation to the images to artificially increase the size of the training dataset and improve the generalizability of the algorithm, as, for instance, flipping the image of an object should not affect its spectrum. Specifically, we flip images horizontally and vertically with a probability of 0.5 and apply random cropping. We do not apply data augmentation to the spectra. We do not apply any data augmentation to the test samples used for evaluation in the following sections.

All models are trained with 2 NVIDIA GeForce RTX 3090s until convergence of the Mean Squared Error (MSE) between generated and ground-truth samples on the validation set. 
We make our code publicly available~\citep{code}.
\section{Evaluation Methods} \label{sec:evaluation}
 
In this section, we present the tools used to evaluate the information content of our artificial spectra. A successful generative model should not only produce galaxy spectra with similar shapes and matching overall features (\emph{i.e.}, a low MSE) compared to their observed counterparts. More importantly, the extracted stellar population properties from the artificial spectra should coincide with those imprinted in the observed spectra. The spectroscopic toolbox we utilize for this contains two stellar population fitting codes (subsection \ref{sec:eval1}) capable of recovering the mean age, metallicity, extinction, and stellar mass of a galaxy from its spectrum. It also encompasses measurements of the strength of prominent emission and absorption lines (Lick Indices) and band-head features (such as Dn4000). A detailed explanation and the corresponding definitions are found in subsection \ref{sec:eval2}. Finally, we assess whether the generative model has learned to link typical AGN emission to photometric features and evaluate it through popular performance metrics in machine learning (subsection \ref{subsec:agn}).

\vspace{-0.9cm} 

\begin{deluxetable*}{cCCCcc}[htb]
\tablecolumns{6}
\tablewidth{15pt}
\tablecaption{Spectral Indices \label{tab:index}}
\tablehead{
\colhead{Index} & \colhead{Blue Side band [$\angstrom$]} & \colhead{Line  [$\angstrom$]} & \colhead{Red side band  [$\angstrom$]} & \colhead{Reference} &  \colhead{Frame} }
\decimalcolnumbers
\startdata
Dn4000 & 3850.000 $–$ 3950.000 &  & 4000.000 $–$ 4100.000 &  \citet{Balogh1999} & air \\
Mg b & 5142.625 $–$ 5161.375 & 5160.125 $–$ 5192.625 & 5191.375 $–$ 5206.375 & \citet{Trager1998} &  air \\
Fe5270 &  5233.150 $–$ 5248.150 & 5245.650 $–$ 5285.650 & 5285.650 $–$ 5318.150 & \citet{Trager1998} &  air \\
Fe5335 & 5304.625 $–$ 5315.875 & 5312.125 $–$ 5352.125 & 5353.375 $–$ 5363.375 & \citet{Trager1998} &  air \\
H$\beta$ &  4827.875 $–$ 4847.87500 & 4847.875 $–$ 4876.625 & 4876.625 $–$ 4891.625 & \citet{Trager1998} &  air \\
\enddata
\tablecomments{SDSS spectra are provided in vacuum wavelengths, but many indices are measured in air. If so, the SDSS wavelengths are converted using \citet{Morton1991}. Yet, the resulting error would be negligible. It should also be mentioned that Dn4000 as a spectral index is measured using the flux per unit frequency ($F_{\nu}$), not flux per unit wavelength ($F_{\lambda}$) as the others. }
\end{deluxetable*}

\subsection{Full spectral fitting} \label{sec:eval1}
A possible way of evaluating the quality of artificially generated spectra is to run full-spectral fitting codes on them. As the name states, these techniques work on the complete wavelength range available, not only on distinct spectral features such as the Balmer lines. Today, they are a standard procedure to determine galaxy properties, including age, stellar metallicity, stellar mass, and dust extinction of composite or spatially-resolved stellar populations. In this work, we used \texttt{FIREFLY}~\citep{Wilkinson2015, Wilkinson2017} and \texttt{pPXF}~\citep{Cappellari2004,Cappellari2017,Cappellari2023}, two well-known non-parametric population synthesis codes available from the astrophysical community. They do not assume a star formation history a priori but try to recover it, along with other properties, from the spectrum and are, therefore, quite general (i.e., concerning merger events). The user should nevertheless not blindly trust the complete star-formation history (SFH); a reduction to a robust young and old population can sometimes be the only way to statistically sound statements \citep{Fernandes2005}. 

\texttt{pPXF} applies a penalized maximum likelihood approach to fit single-burst stellar population models (SSPs) on spectra. By imposing a penalty on pixels that are not well characterized by the templates, it works to minimize template mismatch. One of the advantages of the code is the possibility of simultaneous fitting of gas emission lines along with stellar kinematics (velocity dispersion) and stellar population. With this software, we are using the included model templates from \citep{Vazdekis2016}. We realigned our work to Jupyter Notebook examples\footnote{\url{https://github.com/micappe/ppxf_examples}} available online, which show the use of an additional bootstrapping method \citep{Davidson2008} with \texttt{pPXF} during the fit. As a first step in this procedure, a regularization (smoothing of template weights with \texttt{regul}=10) was applied. The emerging residuals are stored and then used to bootstrap the spectrum $50$ times. This leads to robust average galaxy properties (mean age, mean metallicity, ...) and an estimate for their uncertainties (see also \citealt{Kacharov2018} for more details). \\
When replicating the examples, we strongly recommend using the newer \texttt{.dust-function} for the extinction, not the now obsolete \texttt{.reddening/ .gas-reddening} keywords which are used in the examples. 

\texttt{FIREFLY} is a chi-squared minimization fitting code that fits combinations of SSPs to spectroscopic data, following an iterative best-fitting process controlled by the Bayesian information criterion. This approach has been designed to be a good way to recover galaxy properties, especially in low S/N-regimes, where accurately deriving properties from spectral fits becomes more and more challenging \citep{Goddard2017}. Extinction due to dust is not fitted in a conventional way: A High-Pass Filter (HPF) is used to rectify the continuum before fitting, allowing for the removal of large-scale modes of the spectrum associated with dust and/or poor flux calibration. Regions with nebular emission lines are masked out during the process.\\
The \texttt{FIREFLY} package is supplied with the pre-calculated stellar population models of \citealt{Maraston2011} (MILES \citealt{Falc2011} as the stellar library combined with a Kroupa IMF \citealt{Kroupa2001} in a fuel-consumption approach). We also ran tests with the models used in \citep{Sextl2024} derived with FSPS v3.2 \citep{Conroy2009} and a combination of several stellar catalogs (MILES in combination with additional templates from Post-AGB, WR-stars etc) with a Chabrier IMF \citep{Chabrier2003} and MIST isochrones \citep{Dotter2016, Choi2016, Paxton2011}. This leads to substantially longer run times of the code but not to improved results in the evaluation metrics presented in this section. The lack of hot stars ($T > 9000$ K) in the MILES library \citep{Martins2007} does not seem to play a crucial role in the setup here. We, therefore, remain with the Maraston \& Strömbäck SSPs. The corresponding age grid covers SSPs between 6.5 Myr and the Age of the universe, while the sampled metallicities read $[Z] = -1.3, -0.3, 0.0, $ and $0.3$. Regions with emission lines or absorption features polluted with emission are masked for a functioning fit. As an extinction law, \cite{Calzetti2001} is chosen. 
For the observed spectra, the velocity dispersion ($\sigma_{\ast}$) retrieved from \texttt{pPXF} was used in the input file. These values are in agreement with the error bars with the values for $\sigma_{\ast}$ deposited in the SDSS archive. The artificial spectra tend to show similar velocity dispersions as their real counterparts (see figure \ref{fig:main_hist}). \\
As an additional asset, we also integrate the bootstrapping method from above into the \texttt{FIREFLY} Python routines. Due to longer calculation times, the spectra are only bootstrapped $10$ times. Yet, \texttt{FIREFLY} is a code already designed to work in a low S/N regime, and we will see that the artificial spectra perform equally well with both codes. \\

A key distinction between the codes lies in the treatment of dust: While \texttt{pPXF} assumes a dust reddening law and dust is treated as an adjustable fit parameter, \texttt{FIREFLY} determines the effect of dust prior to the main fitting by comparing the large modes of data and models \citep{Wilkinson2017}. We emphasize that the focus of this work does not lie in the comparison between the different codes or SSPs but in the performance of the predicted spectra in contrast to the true observed spectra. Absolute values from full spectral fitting are always affected by systematic differences in the technique itself and underlying stellar population models and their ingredients \citep{Baldwin2018, Oyarzu2019,Chen2010, Ge2019}. 

\begin{sidewaysfigure*}
    \raggedleft 
    \vspace*{10cm}
        \includegraphics[width=\columnwidth]{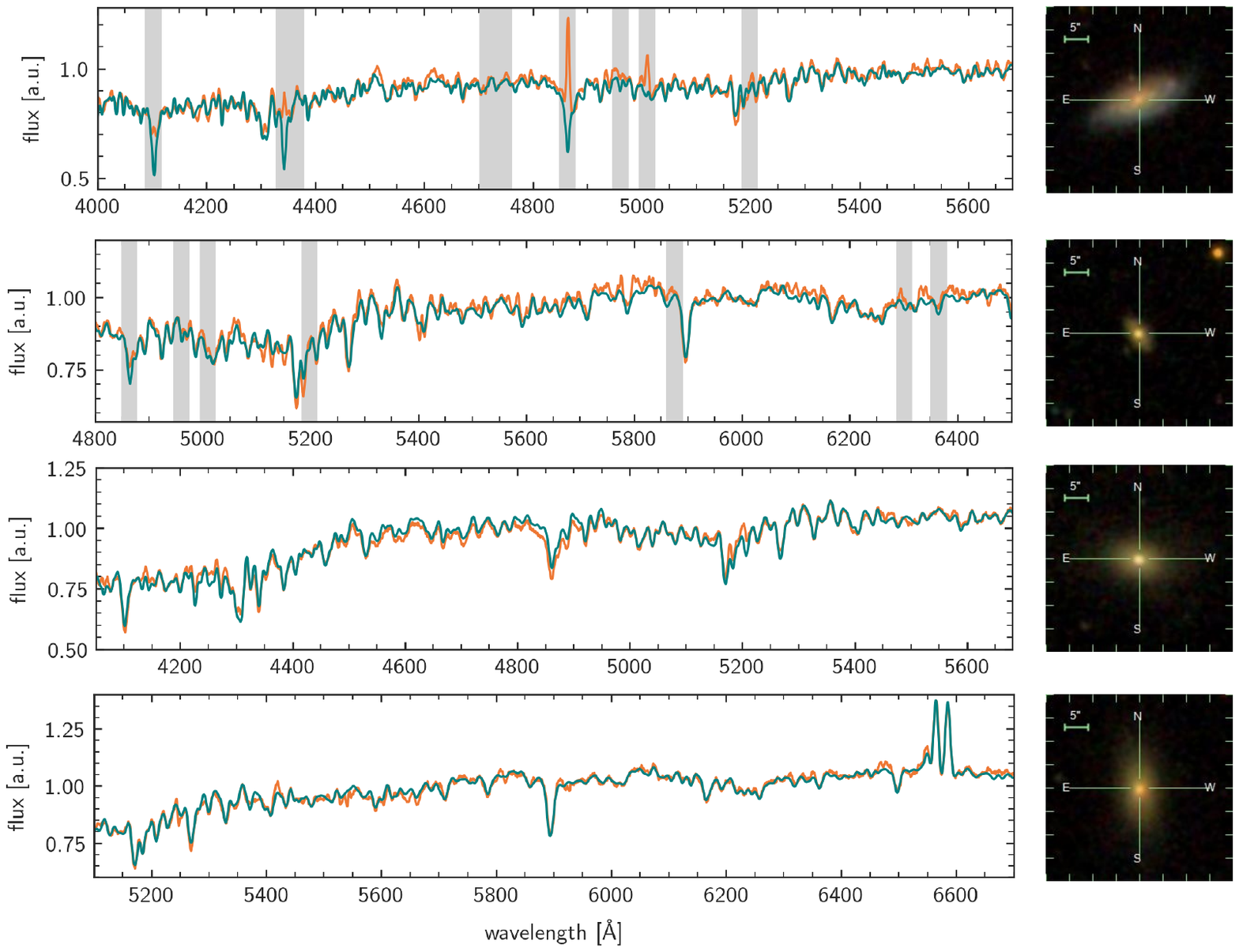}
    \caption{Another four predicted created spectra (orange) and their corresponding full-spectrum fits with a population synthesis code. A \texttt{FIREFLY} result is shown in the first two rows. Regions with potential emission lines (not necessarily visible in each spectrum) are masked out for the fit and displayed here on a gray background. The third and fourth spectra are fitted with \texttt{pPXF} with emission and absorption fitted in parallel. The images on the right show again the corresponding SDSS tri-color images for illustration; the algorithm uses 5-band images. } \label{fig:example_fit}
\end{sidewaysfigure*}

Figure \ref{fig:example_fit} shows four full-spectral fit examples for \texttt{FIREFLY} and \texttt{pPXF} on selected galaxies in the test set.


\subsection{Spectral Features} \label{sec:eval2}

Before the widespread use of full-spectral fitting techniques as described above, a limited number of prominent stellar absorption features were examined. Their strength or equivalent width is not sensitive to flux calibration and is available even at modest resolution and low S/N optical spectra. Extensive research in the 1980/1990s by \citet{Burstein1984, Worthey1994, Worthey1997, Trager1998} led to the Lick index system, a set of 25 optical absorption-line indices, the most commonly used in absorption-line analyses of old stellar populations. Each index in this system is defined by a central ``feature bandpass'' and two adjacent windows, blue- and rewards, for defining so-called ``pseudo-continua'' acting as baselines. Some indices are known to be more age-sensitive (e.g., Hydrogen Balmer lines) or metallicity-sensitive (e.g., Fe, Mg features). The well-known additional dependence on $\alpha$-enhancement ($\alpha / Fe$) complicates the picture, and efforts went into constructing combined metallicity indices with a weak $\alpha$-dependence (see below). \\
In our work, we measure prominent Lick indices (equivalent widths compared to two sidebands) and also Dn4000 as a ``bandhead'' index (difference of two passbands) based on the recipes in the Mapping Nearby Galaxies at Apache Point Observatory (MaNGA) data pipeline \citep{Westfall2019} with the program \texttt{PYPHOT} \citep{Fouesneau2022}. This package does not take into account uncertainties in the flux. SDSS galaxies also do not show strong wavelength-dependent noise, which would distort the measured Lick values \citep{Nersesian2024}. Table \ref{tab:index} shows the utilized indices and their pseudo-continua wavelength ranges. As a robust metallicity indicator independent of $\alpha$-enhancement, the averaged Lick index [MgFe]$^{\prime}$ from \citet{Thomas2003} is computed from our measurements: 
\begin{equation}
    [\textrm{MgFe}]^{\prime}=\sqrt{\textrm{Mg b} \cdot  \left( 0.72 \cdot \textrm{Fe5270} + 0.28 \cdot \textrm{Fe5335} \right) }
\end{equation}
We do deteriorate the spectral resolution to reach the original wavelength-dependent Lick resolution (FWHM[$4000 \, \angstrom$]=11.5 $\angstrom$, FWHM[$6000 \, \angstrom$]=8.9 $\angstrom$). 

Lick Indices were originally mainly used for quiescent galaxies without visible emission lines, but star-forming galaxies in our test sample can have strong H$\beta$-emission. Even the vanishing star formation rates of ellipticals can trigger this emission \citep{Whitaker2013}. In order to prevent contamination in this line index, we measure this index not from the original spectra but from the model fits obtained from \texttt{FIREFLY}. The fit only contains stellar light and no emission from gas. Other spectral features are not affected by this issue.\\
Velocity dispersion broadening has another non-negligible effect on absorption index measurements and has to be accounted for. Many artificial spectra show similar velocity dispersions as the real observations (figure \ref{fig:main_hist} top right), yet some deviate strongly. We apply the method presented in \citet{Zheng2019} to remove the effect of the velocity dispersion altogether, keeping in mind that single SDSS galaxies can show values of $\sigma_{\ast} \approx 400 \, \textrm{km/s}$. The authors assume a simple polynomial relation between spectral indices with and without velocity dispersion broadening and provide suitable coefficients for our used indices:
\begin{equation} \label{eq:sigma}
    x_0=p_0 + p_1 x^{p_2} +  p_3 \sigma^{p_4} +  p_5 x^{p_6} \sigma^{p_7}
\end{equation}
 $\sigma$ is the velocity dispersion measured in $\textrm{km/s}$ and $x$, $x_0$ are spectral index values before/after correction. The values for the coefficients $p_0$ to $p_7$ for each index can be found in \citet{Zheng2019} table 2. The uncertainties traced through this method are usually below $5 \%$. In order to use equation \ref{eq:sigma}, robust velocity dispersions of our artificial and observed spectra are needed. We measure them with \texttt{pPXF} (see \ref{sec:eval1}).


\subsection{Performance on AGN recognition} \label{subsec:agn}

For a typical binary classification task in machine learning, various performance metrics are used to evaluate how accurate the predictions are. In subsection \ref{agn_class}, we will ask the question of whether a galaxy harbors an AGN or not according to its position in the Baldwin-Phillips-Terlevich (BPT) diagram~\citep{Baldwin1981}. For this, the real observed spectra or the artificial spectra are used as a starting point. If both spectra (the true and the predicted one) lead to the same classification as AGN, we speak of a true positive (TP). If the emission lines in both spectra point towards a pure star-forming galaxy, it is a true negative (TN). FN (False negative) denotes the number of AGNs not showing up in the artificial spectra. The three metrics used in our work are:
\begin{align} 
&\textrm{Accuracy} = \dfrac{\textrm{TP} + \textrm{TN}}{\textrm{Total}} \\[0.1cm]
&\textrm{Precision} = \textrm{Purity} = \dfrac{\textrm{TP}}{\textrm{TP} + \textrm{FP}} \\[0.1cm] 
&\textrm{Recall} = \textrm{Sensitivity} = \dfrac{\textrm{TP}}{\textrm{TP} + \textrm{FN}}
\end{align}


An evaluation based on accuracy should only be used in approximately equally sized groups, as unbalanced sizes lead to largely overestimated accuracy scores.

\section{Results} \label{sec:results}

\subsection{Quality assessment}

\begin{figure*}[htb]
        \medskip
	\center  \includegraphics[width=1\textwidth]{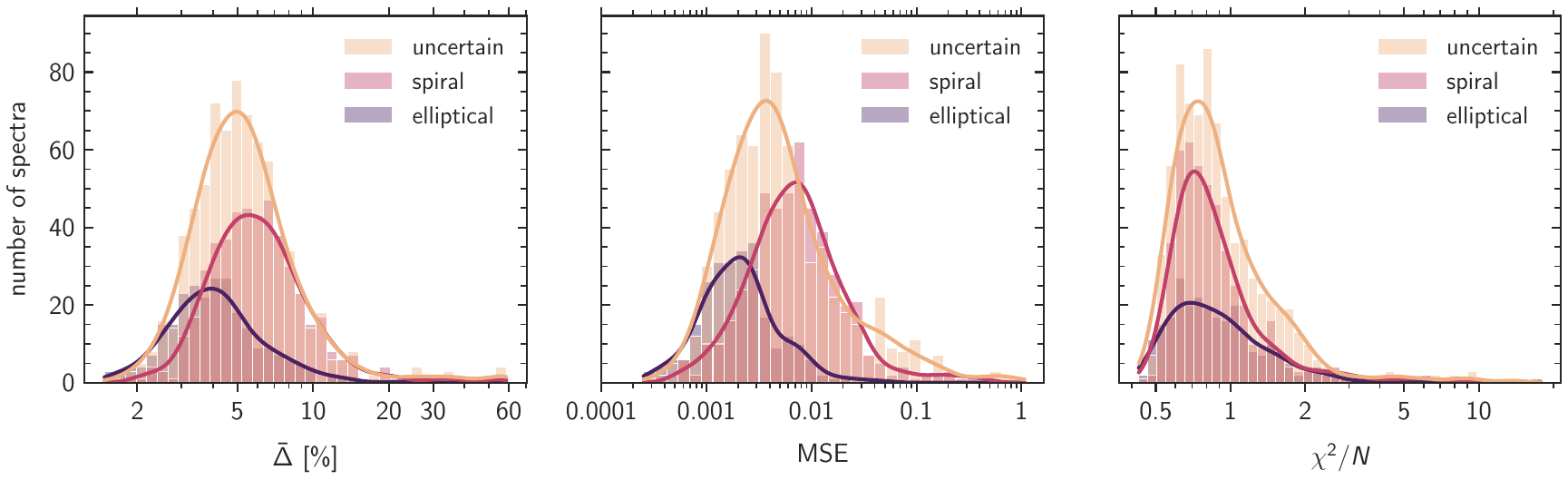}\medskip 
	\caption{Histogram of $\overline{\Delta}$, Mean Square Error (MSE) and $\chi^2/N$ (with N the number of flux-points) of all galaxies in the test set split into different morphologies. When taking into account the uncertainty of the observed spectra, all three groups perform equally well. The morphological categories ``elliptical'', ``spiral'', and ``uncertain'' were determined by the citizen science project Galaxy Zoo \citet{Lintott2008, Lintott2011}). Galaxies were labeled as ``uncertain'' if their images were not clearly voted on as spiral or elliptical. Note that this does not say much about their spectral classification as, for instance, a post-merger starburst galaxy could belong to this group.}   \label{fig:delta_cat}
\end{figure*}

First, we mathematically assess the performance of our generative model before discussing the spectroscopic quality of the predicted spectra. One possible way to evaluate the difference between the predicted and observed spectra is by comparing them at each wavelength point. To do this, we shift the anticipated redshift of the artificial spectra to match the true redshift of the observed spectra. Figure \ref{fig:delta_cat} shows the quality indicator $\overline{\Delta}$  for the whole test set of 1506 galaxies split into different morphological groups defined by the galaxy zoo. It is defined as 

\begin{equation} \label{eq:delta}
    \overline{\Delta} = \dfrac{1}{N}  \sum_{\lambda} \dfrac{|O_{\lambda} - F_{\lambda}|}{O_{\lambda}}
\end{equation}

 with the number of wavelength points $N$, $O_{\lambda}$ and $F_{\lambda}$ represent the observed and the predicted spectra respectively. As such, $\overline{\Delta}$ can be interpreted as a measure of the mean relative deviation between the true and artificial galaxy spectrum (after \citealt{CidFernandes2013}). The median $\overline{\Delta}$ in the test set is 5.5\%, and 90\% of all spectra have values less than 10\%. The inspection of the worst matches ($\overline{\Delta} > 30$)  reveals that the generative model heavily under- or overestimated the strength of the Balmer emission lines in these cases. The strong discrepancy in a few wavelength points affected the overall sum in equation \eqref{eq:delta}; the same holds true for the MSE error. A further inspection indicates that these poorly forecasted galaxies are mostly noisy low-mass starburst galaxies (EW $H \alpha > 35 \angstrom$) categorized as ``spiral'' and ``uncertain'' in roughly equal parts. When calculating $\chi^2=\sum \frac{|O_{\lambda} - F_{\lambda}|^2}{\sigma_{\lambda}^2}$ which incorporates the uncertainty of the observed spectra $\sigma_{\lambda}$, all three morphological groups perform equally well. This is surprising since emission lines are usually narrow in comparison to the width of the broad-band filter, and photons from the continuum, not from the emission line, account for most of the signal. Such a performance can only be achieved by relating emission line fluxes not only to the averaged colors of the galaxy but also to the distribution of colors/magnitudes in the images. How well the generative model recovers line ratios is discussed in section \ref{agn_class}. 
 We do not have error bars for the predicted spectra available as we focused on obtaining a single best estimate for the spectrum of an object. However, while CDMs do not allow us to access the learned distribution directly, one can, in principle, sample multiple candidate spectra per image. In future work, we plan to explore how to use these to provide uncertainty estimates for our predicted spectra.

\subsection{Redshift estimate}

\begin{figure}[htb]
\vspace{0.3cm}
\center \includegraphics[width=\linewidth]{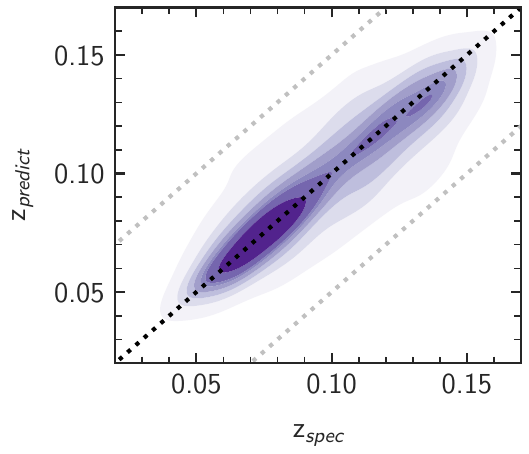}\medskip 
	\caption{Comparison between the spectroscopic redshift from the observed spectra (z$_{\textrm{spec}}$) and the prediction from our generative AI for the galaxies (z$_{\textrm{predict}}$) in the test set. Instead of single data points, the overall kernel density estimate is shown for a clearer point of view. The grey dotted lines mark the regions with catastrophic outliers with $|\Delta z /(1+z_{spec})| >0.05$. The black dotted line shows the one-to-one relation. } \label{fig:zpredict}
\end{figure}

Since we do not remove the redshift in our galaxy sample, the generative model also predicts not only the spectrum itself but also an accompanying redshift. We measure this ``photo-z'' in the artificial spectra and compare it to the true spectral redshift (z$_{\textrm{spec}}$). This term is slightly odd as we measure the redshift in a spectrum, but this underlying artificial spectrum is a prediction on photometric images alone. As there are no suitable words to describe this situation, we refer to these redshift estimates as ``predict-z''. Often used statistics concerning redshifts are:
\begin{itemize}
    \item $\Delta z$ = (z$_{\textrm{pred}}$-z$_{\textrm{spec}}$)/(1+z$_{\textrm{spec}}$)
    \item $\sigma_{MAD}=1.4826 \cdot |\Delta z - \textrm{median}(\Delta z)|$
\end{itemize}
The MAD value (for Median Absolute Deviation) is a common tool for comparing the quality of the predicted redshifts. It is a general measure of dispersion similar to the standard deviation but more robust to outliers. \citet{Pasquet2019}, whose training set we are using, achieved a $\sigma_{MAD} = 0.00912$, signiﬁcantly lower than other machine learning techniques on the same samples. Later work with various ML architectures expanded this work, e.g., \citet{Henghes2022,Schuldt2021,Lima2022}.
On our rather limited redshift sample $(0.05 < z < 0.15)$ we achieve values of $\sigma_{MAD} = 0.01177$ and obtain 2 from $1\,506$ outliers with $\Delta z >0.05$, see Figure~\ref{fig:zpredict}. Unlike other methods, this competitive result is only a byproduct of our generative AI, which is not limited to predicting redshift alone. However, we tested here only on a relatively narrow redshift range. The cited literature mostly focuses on redshifts up to $z \sim 1$ or beyond. 

Predicting redshift from images, in general, offers a significant advantage over traditional methods that rely on limited and biased information (i.e., only colors and magnitudes) taken from pre-processed catalogs \citep{Isanto2018}. The user is simply not biased in selecting measured properties obtained from the galaxy image beforehand. The question of which features are most important does not present itself when using images as basically all features imaginable are present \citet{Hoyle2016}. While supervised convolutional neural networks are the natural choice for redshift prediction with photometric images \citep{Brescia2021}, we show that conditional diffusion models are also successful - when a small detour over spectra is taken. Improving photometric redshift estimates is one of the most pressing needs in the next generation of photometric surveys to unlock their full potential \citep{Newman2022}. Further work will show whether our approach can help in this regard.

\subsection{Spectral Indices}

\begin{figure*}[htb]
\gridline{\fig{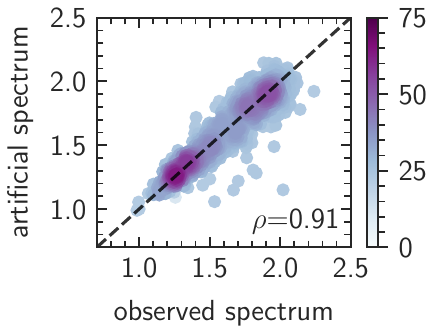}{0.32\textwidth}{(a) Dn4000 Index} \vspace{-0.1cm}
          \fig{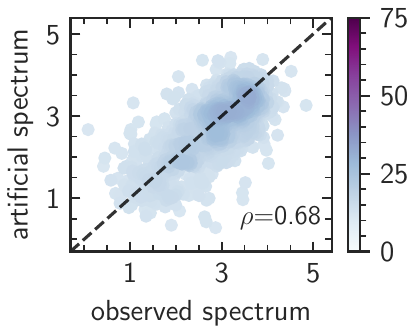}{0.305\textwidth}{(b) [MgFe]$^\prime$ Index} \vspace{-0.1cm} \fig{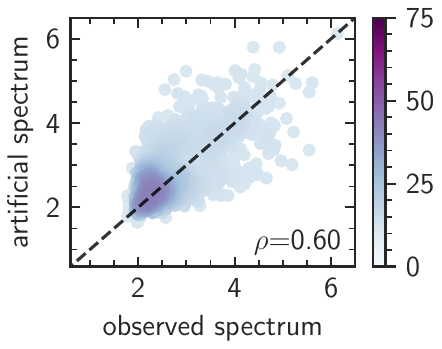}{0.31\textwidth}{(c) H$\beta$ Index } } 
          
\gridline{  \fig{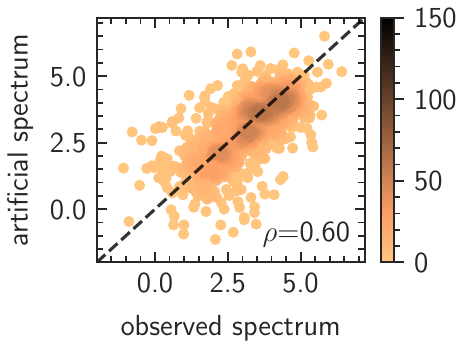}{0.321\textwidth}{(d) Mg b Index}  \vspace{-0.1cm}
            \fig{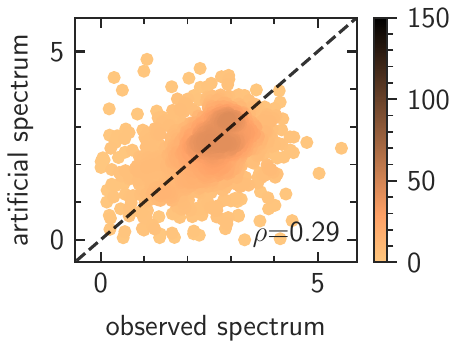}{0.321\textwidth}{(e) Fe5270 Index}  \vspace{-0.1cm}
            \fig{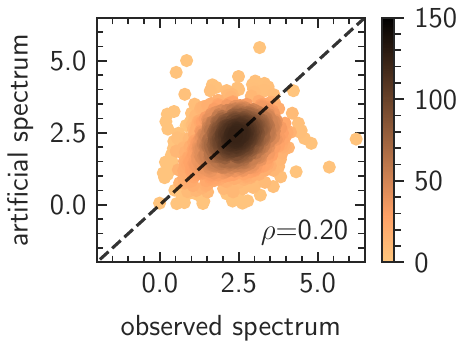}{0.321\textwidth}{(f) Fe5335 Index}  }
            
\caption{Comparison of Lick Indices measured using the observed spectra and the predicted spectra. Spiral galaxies were excluded. The color coding of the density plots indicates how many galaxies reside at a specific (x,y)-position and is kept frozen in each row for comparability. Pearson correlation coefficients $\rho$ are printed in each subfigure. Generally, the coefficient’s value ranges from +1 (perfect positive correlation) to -1 (perfect negative correlation), with 0 indicating no correlation.  \label{Lick_Index}}
\end{figure*}

In the previous section, we showed that the observed and predicted spectra agree quantitatively to a high degree. To further quantify the ML output, we measure spectral indices at several key absorption features in all galaxies in the test set. These can be sensitive to metallicity (Fe5270, Fe5335, Mg b, [MgFe]$^\prime$), age (H$\beta$) or both (Dn4000, \citealt{Afonso2020}).
For proper measurement of the correct spectral regions, all spectra are shifted into the rest frame beforehand. Figure \ref{Lick_Index} shows the velocity dispersion corrected indices for the predicted and the observed spectra of the uncertain \& elliptical morphological group. 

For [MgFe]$^\prime$ and Dn4000, we see a strong correlation between observed and predicted values, which also translates into high Pearson correlation coefficients ($\rho=0.71$ and $\rho=0.91$). 
For Mg b, the values correlate well, $\rho=0.60$, yet the correlation is improved heavily up to $\rho=0.78$ if only the subset of ellipticals is used. \citet{Buzzoni2015} argue that the forbidden [N I] line emission at $5199 \, \angstrom$  can add to the red Mg b pseudo-continuum and enlarge the index value (and lead to higher metallicities). Since [N I] is generally correlated with [O III] emission, Mg b values in star-forming galaxies are generally unreliable. A wrong prediction of [N I] from our generative model might, therefore, negatively influence the comparison to the observed spectra. \\
For the two iron indices, a one-to-one relation is no longer clearly identifiable. The correlation coefficients are below $0.5$ for all indices; Fe5335 is as low as $\rho = 0.2$. These two neighboring features are generally less prominent in the galaxy spectra. Fe5270, in particular, testifies of cool main sequence stars (around 4500 K) and MK III giants \citep{Buzzoni2009}. The fraction of light from such stars in our spectra is small in comparison to the contribution of luminous main-sequence or supergiant stars.

\subsection{Full-spectral fitting results}

\begin{figure*}[htb]
	\center \includegraphics[width=\textwidth]{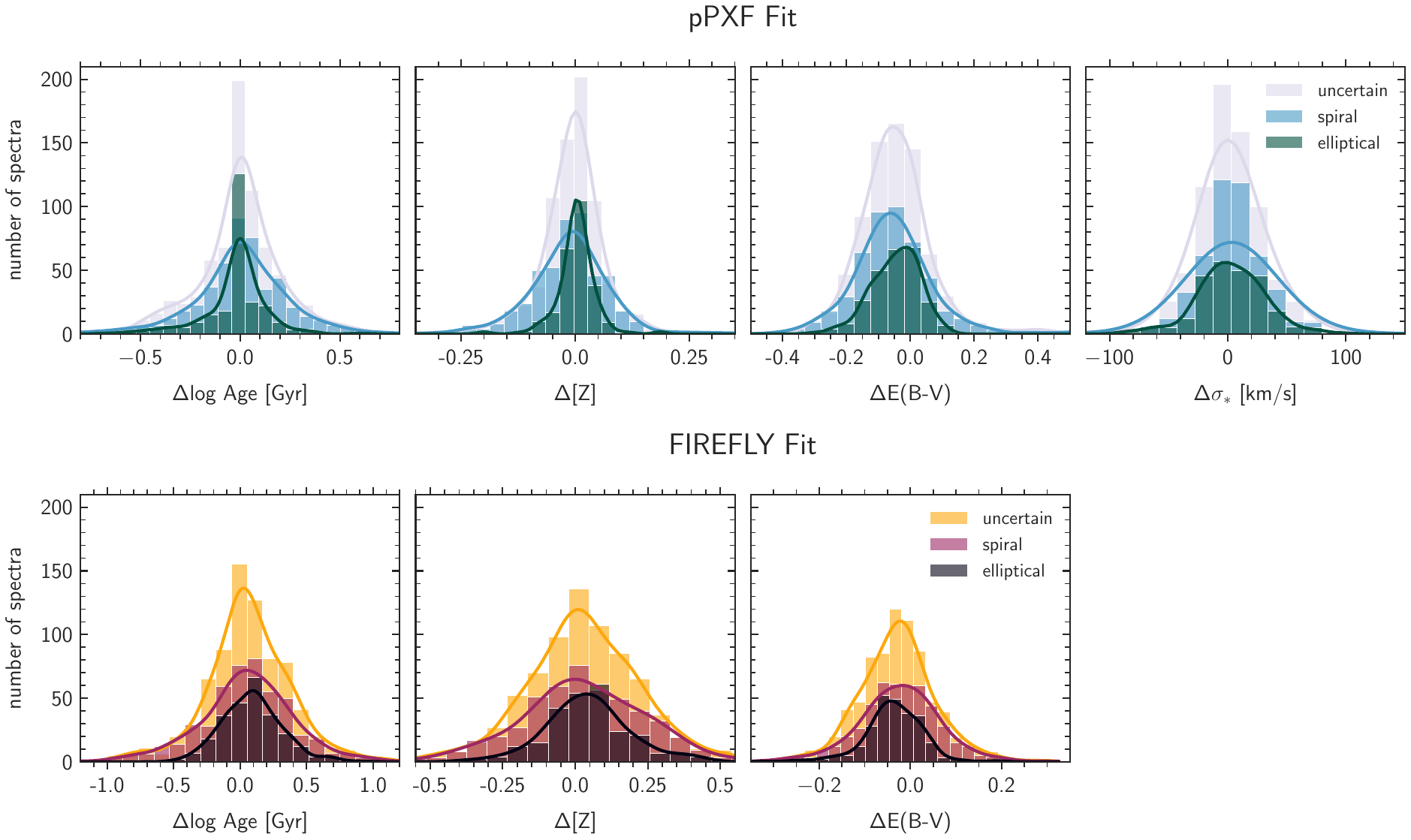}\medskip 
	\caption{Distribution of differences (observed minus predicted) in the derived physical quantities for the used fitting codes. Each histogram shows the difference in inferred quantities (age, metallicity, E(B\nobreakdash-V), $\sigma_{\ast}$) between the real and predicted spectra. The first row in green colors shows the results for the fitting code \texttt{pPXF}, and the second row shows the results for FIREFLY. Different morphological categories as marked before. } \label{fig:main_hist}
\end{figure*}

The metallicity of stars in galaxies is an important indicator of the chemical evolution of a galaxy. Yet, measuring gas or stellar metallicity from photometric data alone presents challenges due to the age-metallicity-dust degeneracy. This means that a galaxy can appear red for various reasons, including the cessation of star formation, high metallicity, or strong attenuation. Additional information like morphological features can help to lift these problems. In the second column of figure \ref{fig:main_hist}, the differences between metallicity derived for the observed spectra are compared with the values determined by the predicted spectra. The first row shows \texttt{pPXF} results, the second row shows the FIREFLY fits. 

\begin{figure}[htb]
	\center \includegraphics[width=1\columnwidth]{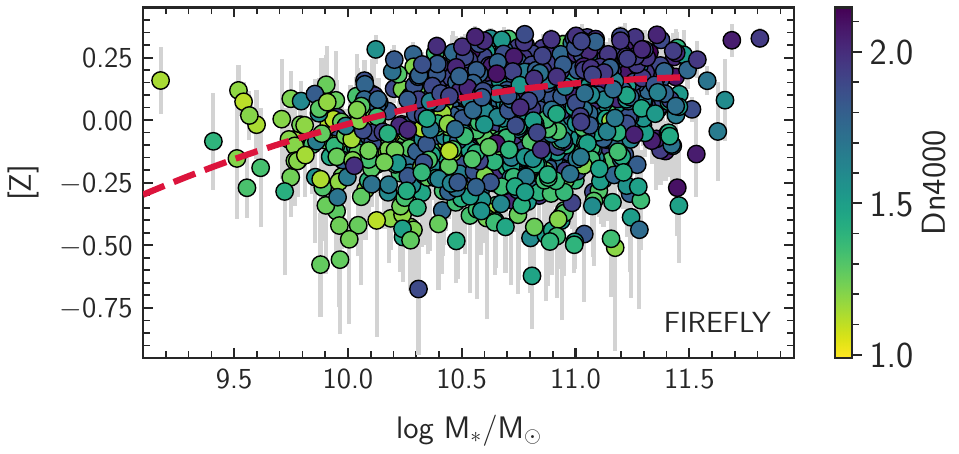}\medskip 
    \center \includegraphics[width=1\columnwidth]{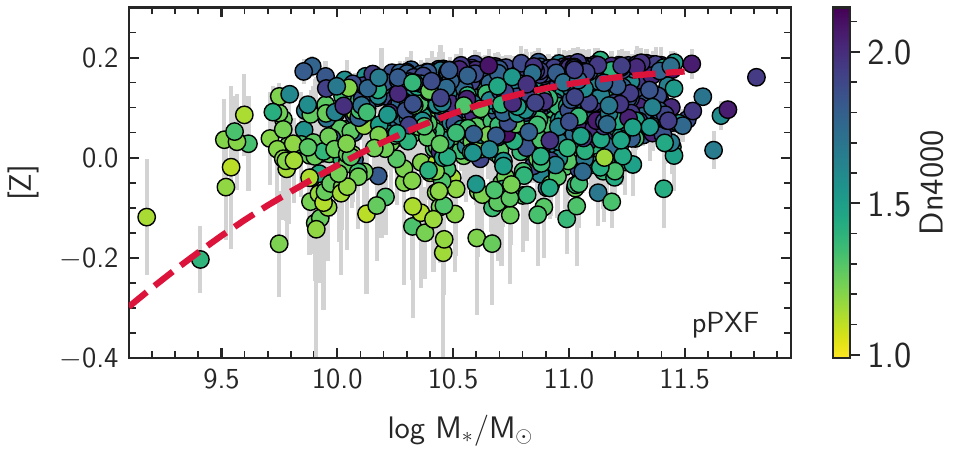}\medskip
	\caption{The final mass-metallicity relationship for the predicted SDSS spectra. The result for \texttt{FIREFLY} can be found in the upper panel, and for \texttt{pPXF} in the lower one. Error bars in the metallicity indicate the 1-sigma uncertainty derived by the corresponding bootstrapping procedure. The stellar mass estimates come from the SDSS photometry alone \citep{Maraston2009}. The dashed red lines show the galaxy lookback evolution models from \citet{Kudritzki2021,Kudritzki2021c}. The colors indicate the value of Dn4000 for each galaxy. Low values (in yellow) point toward young galaxies, and high values (dark blue) towards older stellar populations. SDSS galaxies mostly occupy the space of old and metal-rich populations at low redshift; we also find this in the test set.}     \label{fig:mass_metal}
\end{figure}

Concerning \texttt{pPXF}, 86\% of all predicted spectra coincide in metallicity within 0.10 dex. Taking into account the obtained error bars from the bootstrapping procedure, which are of the order of 0.07 dex in the median, 95\% of metallicity values derived from the predicted spectra coincide with their real counterparts. As expected, the elliptical galaxies show narrower distributions than spirals as they have an overall narrower range of metallicities (see for instance \citealt{Li2018}, figure 5) and easier star-formation histories. For FIREFLY, the scatter is larger. 44\% of all galaxies coincide within 0.10 dex and 70\% within 0.2 dex. Higher uncertainties in the \texttt{FIREFLY} metallicity (0.11 dex in the median) due to the lower number of bootstrapping cycles do not counterbalance this. The values of the flux-weighted age show similar results. For \texttt{pPXF}, the scatter is 0.23 dex in the median, and for \texttt{FIREFLY}, 0.33 dex. For that, the deviations in reddening are notably smaller for FIREFLY independent of the morphological group (0.08 dex vs 0.13 dex in \texttt{pPXF}). This most likely has to do with the different fitting algorithms themselves, which are further explored in the next subsection. The mean of $\Delta$E(B-V) is not exactly at zero, but at $\sim-0.05$ which means that the GenAI predicts dust extinction values slightly too high. This effect is consistently found in all galaxy types as well as fitting codes, also in the larger test set found in the appendix \ref{app:size}. \\
The velocity dispersion in the spectra cannot be fitted with FIREFLY (see section \ref{sec:eval1}); this is only implemented in \texttt{pPXF}. The shown distribution in the top right of figure \ref{fig:main_hist} has an overall mean of -1.4 km/s and a standard deviation of 40 km/s. The distribution is visibly smaller for ellipticals (in dark green), with a standard deviation of 29 km/s. Whether these numbers are good enough for a scientific application depends on the task in mind. But one should not forget that precise measurement of stellar velocity dispersion provides additional insights into the gravitational potential well that encompasses the stars. Since this potential is primarily influenced by dark matter, the velocity dispersion also indirectly indicates the characteristics of the dark matter halo \citep{Zahid2016}. Even though this quantity is of prime scientific importance, the authors are unaware of attempts in the literature to predict the stellar velocity dispersion from photometry. \\
In the closing of this chapter, the absolute values of the metallicity can be considered. In figure \ref{fig:mass_metal}, the overall mass-metallicity relationship is shown for the predicted spectra (top with FIREFLY, bottom with \texttt{ppXF}). These coincide overall with the lookback evolution models by \cite{Kudritzki2021}. Assuming a redshift-dependent relationship between gas mass and stellar mass enabled them to derive numerical models of chemical evolution that are easy to calculate. The all-in-all scatter in the mass-metallicity relation is expected as it also depends on SFR as a third parameter \citep{Mannucci2010}. Galaxies with lower SFR tend to have higher metallicities at the same stellar mass and vice versa. 
There is a slight discrepancy between the values derived from the two different codes and SSPs. Metallicities with \texttt{pPXF} tend to be systematically lower by $ \sim 0.1$ dex. This was also observed in \citet{Oyarzu2019} fitting spatially resolved passive early-type galaxies from the MaNGA survey. One possible origin might be the spacing of the metallicity grid of the SSPs or the overall spectral library.

\subsection{Degeneracies Between Age, Metallicity and Reddening}

It has been long recognized that optical spectra exhibit degeneracies in terms of age, metallicity, and dust properties. This means that different stellar populations with varying ages, metallicities, and dust properties can have nearly identical optical spectra, making it challenging to distinguish between them based solely on their observations \citep{Worthey1994}. With the advent of full-spectral fitting, this problem has become more pressing. The breakdown of the integrated spectrum into a combination of different building blocks (the simple stellar populations) does not necessarily lead to a unique solution, meaning that a different set of SSPs can also create the same flux output. Also, slight fluctuations due to noise can impact the fitting result obtained. There exist intrinsic limitations to the precision to which age and metallicity can be determined without reformulating the problem (i.e., by having a larger wavelength coverage \citealt{Lopez2016}, or using higher S/N data). 

\begin{figure}[htb]
	\center \includegraphics[width=1\columnwidth]{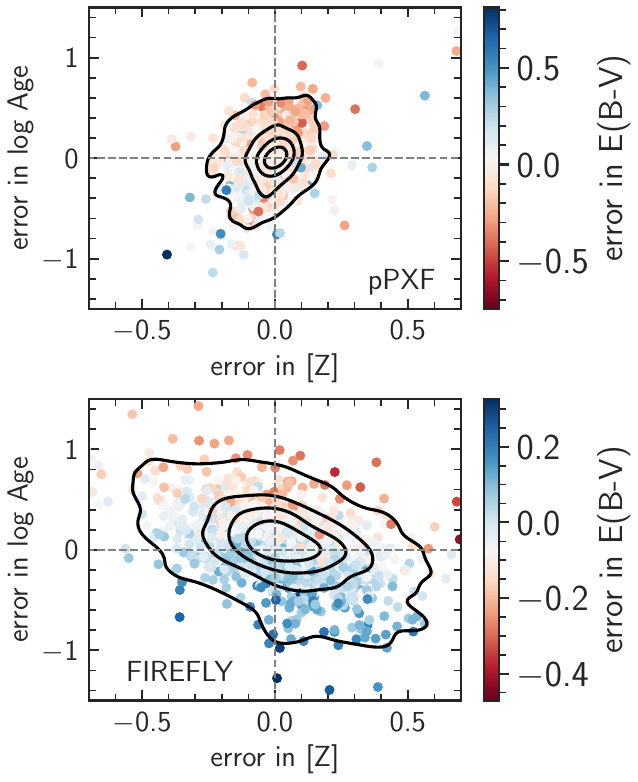}\medskip 
	\caption{Visualization of the degeneracy for both codes. The x-axis (y-axis) depicts the difference between the metallicity (log Age) from the observed and predicted spectrum. In other words, error in $[Z]$ means $[Z]_{obs}-[Z]_{pred}$. The color-coding shows the error in the reddening. Contours are again drawn with a kernel estimate representing a 2D histogram analog. Whereas FIREFLY incorporates an expected age-metallicity degeneracy, \texttt{pPXF} shifted the problem more towards a blurring in age-reddening.}  \label{fig:degeneracy}
\end{figure}

Fitting the rather low S/N SDSS galaxy spectra is also expected to be plagued by this problem. We, therefore, use \texttt{pPXF}, which uses a regularization procedure that is said to be a suitable mathematical tool for such an ill-posed problem. Additionally, we implemented a bootstrapping procedure that tackles the impact of noise in the spectra. Yet, both codes still show degeneracies in age, metallicity, and reddening, which is not surprising. Figure \ref{fig:degeneracy} shows the difference of metallicity between the predicted and observed metallicity on the x-axis and the deviation in log age on the y-axis. The overall color coding marks the discrepancy in the color excess E(B\nobreakdash-V), i.e., reddening. FIREFLY (bottom panel) shows a preferred degeneracy in age and metallicity, whereas, for \texttt{pPXF}, a degeneracy between age and color excess emerges. This is identical to what \citet{Woo2024} found using mock-spectra from the magnetohydrodynamical simulation IllustrisTNG\footnote{https://www.tng-project.org/}. Yet, we discourage a direct comparison of both codes as they do not use identical SSPs in our case. Nevertheless, the generative AI's predicted spectra can be used with both codes without problems despite incorporating completely different fitting procedures and templates. Also, it seems that a not negligible part of the deviation in the physical quantities between the predicted and observed spectra seen in the histograms of Figure \ref{fig:main_hist} do not necessarily come from a different information content of the observed/predicted spectral pair but from the overall degeneracies which plague full-spectral fitting overall in this S/N regime. As a result, even better predictions of the generative AI might not inevitably lead to better conformity as the limiting factor is the reconstruction of physical quantities from spectra, not the spectra themselves.

\subsection{Bimodality of Galaxies}

\begin{figure*}[htb]

\includegraphics[width=1.07\linewidth]{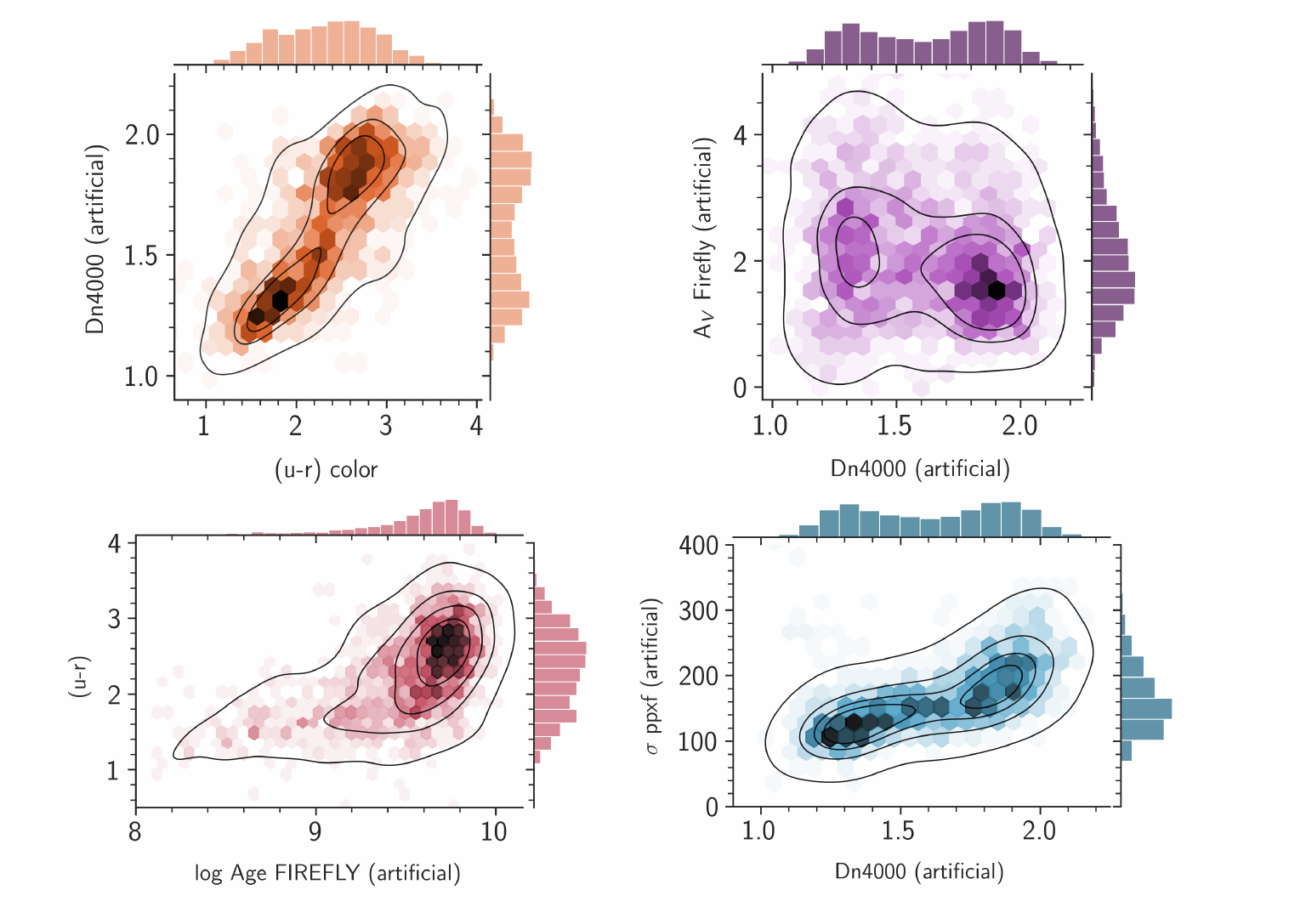}
	\caption{Bimodality of the galaxy population in the test sample. All shown quantities are retrieved exclusively with artificial spectra and photometry. No observed SDSS spectra from the test set were used. The color coding in each subplot indicates again how many galaxies reside at a specific (x,y)-position; darker means more galaxies. Additionally, kernel density estimates (KDE) are overdrawn to highlight the bimodal behavior in mean stellar age, extinction, velocity dispersion, extinction, and color. \label{bimodality}}   
\end{figure*}

A further testing ground for the predicted galaxy spectra is the bimodality of galaxies. 
For decades, a separation of the galaxy population into two distinct groups has been observed in various physical quantities, including color, mass, age, and spectral indices such as Dn4000. This bimodality suggests that galaxies can be broadly classified into two categories: one population dominated by older, more massive, and redder galaxies with lower star formation rates ('the red sequence') and another population consisting of younger, less massive, bluer galaxies with higher star formation rates ('blue cloud') \citep{Holmberg1958, Roberts1994, Strateva2001}. The analysis of the predicted spectra is able to reproduce this two-fold distribution. Figure \ref{bimodality} shows the relation of physical quantities (reddening, stellar age, $\sigma_{\ast}$, Dn4000, u-r) in relation to each other. All of these data points were solely derived from the artificial spectra and (u-r) color from photometry. As most of these relations are already seen partially in photometric colors itself \citep{Baldry2004}, our machine learning algorithm is supposed to pick them up, and this subsection can therefore be seen as a consistency check.

The relations in the upper panels and the lower left in figure \ref{bimodality} were shown in an identical manner for SDSS galaxies in the work of \citet{Mateus2006} using the full-spectral fitting code \texttt{STARLIGHT} \citep{Fernandes2005, Fernandes2007}. Their analysis also goes into detail about why one can expect these relations. On this occasion, it should be noted that the depicted log Age (mean stellar age), is actually the mean flux-weighted stellar age, i.e. a quantity biased towards the age of stars that produce most of the flux in a spectrum. These are mostly very young, high-mass stars formed in recent star formation episodes, which die off quickly (on the scale of less than several hundred million years). Dn4000 and colors are also heavily affected by the flux of these types of stars, which drives some of the observed bimodal behavior seen in Figure \ref{bimodality}. \\
Yet, \citet{Mateus2006} used the original available SDSS spectra for their analysis at that time (DR2). We derived the same relations from purely AI-predicted spectra. This reinforces the idea that the generative AI not only produces spectra that have an overall suitable shape but also captures the nature of the galaxies. Morphology, size, and colors of the galaxy are interlinked with the properties of the stellar populations through complicated relations \citep{Skibba2009,Darg2010}. Our method is capable of taking full advantage of this. \\
The most interesting relation not mentioned yet is the relation in the lower right panel of figure \ref{bimodality}. With \texttt{pPXF}, we are able to retrieve values for the velocity dispersion $\sigma_{\ast}$ of the predicted spectra. So, at the end of the day, our generative AI predicts $\sigma_{\ast}$ from photometric broad-band images alone. In the literature, $\sigma_{\ast}$ is commonly measured with the help of spectra or indirect with the help of scaling relations \citep{Bezanson2012}. In practice this can mean for instance that the stellar mass of early-type galaxies (ETGs) is measured photometrically and in a second step $\sigma_{\ast}$ is inferred from its correlation to stellar mass. Here, we derive the values through the utilization of \texttt{pPXF} on the artificial spectra, but other template-fitting techniques are also possible. When taking the complete test set (no split in different groups), the measurements of $\sigma_{\ast}$ of the observed and the predicted spectra agree with a negligible median offset of 0.5\% with a dispersion of 20\%. There seem to be no other attempts (machine-learning or traditional) on the prediction of $\sigma_{\ast}$ in the literature from photometric images. As the colors of a galaxy are not affected by a different velocity dispersion given a spectrum \citep{Tortorelli2024}, the 2D information seems to be the deciding factor here. The generative AI seems to pick up the fundamental plane of elliptical galaxies (relation between the effective radius, average surface brightness, and central velocity dispersion) and is therefore able to predict $\sigma_{\ast}$ values for the corresponding galaxies. For late-type galaxies (LTGs) the situation is less clear, but even there relations between $\sigma_{\ast}$ and other physical properties can be found \citep{Napolitano2020}. But this also explains why the generative AI is doing a better prediction task for elliptical galaxies (see again Figure \ref{fig:main_hist} top right).

\subsection{AGN classification} \label{agn_class}

\begin{figure*}[htb]
\includegraphics[width=.48\linewidth]{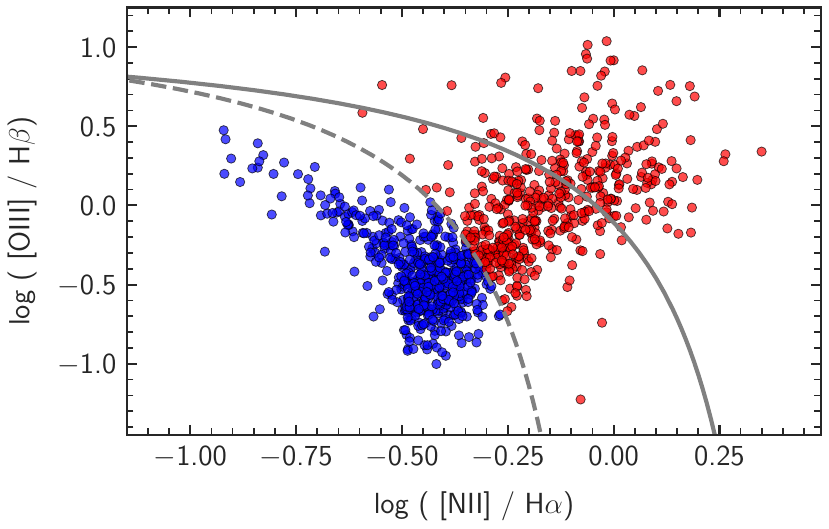}\hfill
\includegraphics[width=.48\linewidth]{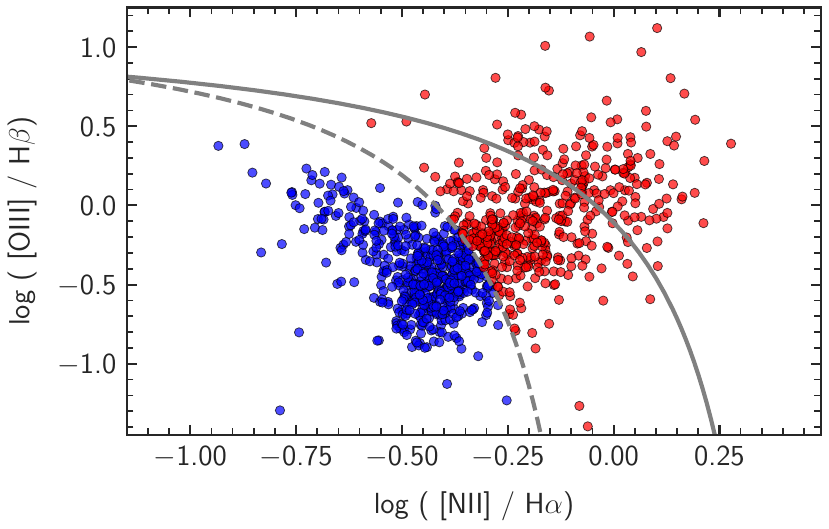}
\caption{BPT diagram for the real spectra (left) and the predicted spectra (right). Pure star-forming and possible AGNs are split in both cases by the empirical relation found by \cite{Kauffmann2003}  (dashed grey). The additional solid grey line from \citep{Kewley2001} is drawn as an additional guideline. Galaxies in the region between the two lines (often labeled as composites) probably contain AGN but are still likely to be dominated by the contribution of star formation.}     \label{fig:BPT}
\end{figure*}

As a final quality test, we focus on AGN recognition in emission-line galaxies. As the generative model predicts the complete spectrum of the galaxy, not only stellar light, it suggests itself to make use of the predicted emission lines. We make use of the classic BPT diagnostic diagram \citep{Baldwin1981}, which compares the relative strength of collision lines of metals to recombination lines of hydrogen. The primary diagram assesses the line ratio [O III]$ \lambda 5007$/ H$\beta $ to [N II] $\lambda 6584$/ H$\alpha$. Due to the pairs of wavelengths being close together, these ratios are not affected by differential extinction. We measure the lines after subtracting the fit of the stellar population from FIREFLY. Figure \ref{fig:BPT} top shows the BPT diagram for the observed spectra. The galaxies tend to lie in two well-defined sequences in the BPT diagram, leading to a characteristic ``seagull'' shape. The left-wing sequence in dark blue is associated with star-forming galaxies, while the right-wing sequence in red is associated (partially) with other ionization mechanisms (mostly AGN). The solid line represents the classification curve from \citep{Kewley2001} defining the upper limit for finding pure SF galaxies. Galaxies were labeled as passive when the equivalent widths (EW) of H$\alpha$ were smaller than 1 $\angstrom$ or when a visual examination showed no spectra left after the subtraction of the continuum. This threshold has the advantage of being independent of any S/N ratio, which is not available in predicted spectra in the first place. \cite{Fernandes2011} argue that this criterion is independent of data quality and has a better astrophysical meaning. We measure the EW of H$\alpha$ with the python package \texttt{specutils}. More problematic is the case when one of the four prominent emission lines used in the classical BPT diagram is missing, as this galaxy cannot be placed in figure \ref{fig:BPT}. Our simple solution is then the assignment as a passive galaxy even though this is not physically correct: The center of this galaxy can simply be (partially) dust-obscured. Ratios between obscured and un-obscured AGN can even be as large as $\sim 3$ \citep{Ballantyne2011}. As this is a general problem of the optical bands we cannot solve, wavelength regimes in the IR provide useful alternatives with less extinction and open up a realm for new line diagnostics \citep{Pentericci2023}.  

\begin{table}[htb]
\centering
\begin{tabular}{||cc|ccc|c||}  
\hline \hline
        &          &          &  predicted   &        &      \\
        &          & passive & pure SF  & AGN         &      \\ \hline
        & passive  & 359    & 33        & 72           & 464  \\  
 actual & pure SF  & 23     & 448       & 68          & 539  \\
        & AGN      & 45      & 88        & 370         & 503  \\ \hline  
        &          & 427    & 569       & 510        & 1506 \\ \hline \hline
\end{tabular}
\caption{Confusion matrix of the test set: The whole test set contains $1\,506$ galaxies. Each row of the table indicates the actual class and each column represents the predicted class label. A value in the cell is a count of the number of predictions made for a class that are actually for a given class. The diagonal elements count correct predictions, whereas the off-diagonal elements count the galaxies where the emission line ratios of the predicted spectra suggest a wrong class. \label{tab:table-AGN}}
\end{table}

The BPT diagram of artificial spectra is shown in the right panel of Figure \ref{fig:BPT}. From the $1\,506$ synthetic spectra, $1\,079$ also showed measurable emission according to the criteria from above. Table \ref{tab:table-AGN} shows the $3 \times 3$ confusion matrix, with rows corresponding to actually being AGN, SF, and passive and columns corresponding to the predicted classification from the artificial spectra. In other words, each row of the matrix represents the objects in the actual class (passive, pure SF, AGN), while each column represents the galaxies in the predicted class from the artificial spectra. From these values, the evaluation metrics can be calculated. The accuracy of predicting a possible AGN is 82\%; the precision resides at 73\%. The recall is on the same level with 74\%. Usually, precision (everything flagged AGN should indeed be an AGN) and recall (not missing any AGNs) are in tension; tuning machine learning on one of these quantities usually diminishes the other, and the optimal trade-off depends on the task at hand. As this paper acts as a feasibility study, the obtained performance metrics already show encouraging results, especially as these AGN predictions come for free from the generated spectra. No specific algorithm solely used for this purpose had to be trained. For instance, \citet{Cavuoti2014} tried to answer the same question (distinguishing AGN and non-AGN with photometry) using various machine-learning methods applied specifically for this task. Instead of images, they used dereddened SDSS colors, the dereddened magnitude in the r band, the fiber magnitude in the r band, and the photometric redshift. A support vector machine (SVM) achieved the best result with an accuracy of 76\% with a test set size of 25466 objects. 

For a proper analysis, the uncertainty in the measurements has to be taken into account. The classifications for our observed spectra will certainly not be completely correct and contain some errors. We follow \cite{Richardson2016} in estimating the error budget: H$\beta $ and [O III] $\lambda 5007$ are free from overlaps with other prominent emission lines and, therefore, rather easy to fit with a Gaussian. Uncertainties arise through noise in the flux and mainly systematic calibration errors, with an estimate of 15\%. Contrarily, H$\alpha$ is partially blended with [N II] $\lambda \lambda 6548, 6584$, which leads to higher errors around $20\%$ for H$\alpha$ and 30\% of [N II] $\lambda 6584$. Measuring the equivalent width of H$\alpha$ was multiplied with a Gaussian error with a standard deviation of 1 to incorporate uncertainty. Some measured EWs are slightly lower than 1 $\angstrom$, and the corresponding galaxy is labeled as passive; with the error budget, they can now find their way in the diagram. All of these errors can be propagated and give a rough probability for the label of the galaxy from the observed spectra (i.e., 95\% probability of being an AGN). The same analysis can be done for the artificial galaxy spectra. Now, random number draws can be drawn where each label (true and predicted) is flipped according to the assigned probability \citep{adebayo2023quantifying}. Galaxies in the upper right corner of the BPT diagram have low changes of classification other than AGN (the BPT diagram is in log-scale), whereas pure star-forming galaxies in the vicinity of the Kauffmann line can swap more easily towards AGN candidates. We run $5000$ experiments with a simple random number generator in Python and obtain the following uncertainties for the evaluation metrics (see subsection \ref{subsec:agn} again for their definitions) from the slightly changed labels: 

\begin{itemize}
    \item  Recall = Completeness =  0.736 $\pm$ 0.018
    \item  Precision = 0.725 $\pm$ 0.016
    \item  Accuracy = 0.818 $\pm$ 0.011
\end{itemize}

As the BPT diagram is on a log scale, uncertainties in the evaluation metrics turn out to be quite small in the end. We still show them here for completeness. More interesting is the reason why the generative AI can predict correct emission lines in the first place. This is at first puzzling since most of the photons in a given broad-band filter come from the stellar continuum and not from an emission line except in extreme cases. Only the broadband photometry of infant massive star clusters is known to be heavily affected by nebular emission, both in lines and continuum \citep{Reines2010}. Yet, our algorithm produces the correct AGN-emission line features only having the photometric images at hand. On second thought, AGNs do not occur at random. A large-scale bar, easily identifiable in photometric images, can, for instance, fuel infalling gas into the central regions of a galaxy (if gas is available in the first place) and trigger both central star formation and the activation of AGNs. Also, star formation activity is stronger in galaxies with large bars \citep{Oh2012}. Work on spatially-resolved spectra in the CALIFA survey by \citet{Lacerda2020} showed further links between AGNs and their hosts. AGNs seem to be concentrated in the high-mass, high-metallicity regime. Their ages are between those of pure blue cloud and red sequence galaxies, and some morphological types are preferred (Sab-Sb types). Additionally, galaxy color and morphology can be affected by the presence of an AGN \citep{Pierce2010}. All of these relations, some of them stronger than others, help relate the optical emission lines with the photometric images. Therefore, it is no surprise that photometric colors (solely colors) correlate with the equivalent widths of emission lines \citep{Abdalla2008}. Yet these authors also found a certain degree of degeneracy between the colors of many passive galaxies and SF ones. The spatial 2D information of colors we provide seems to break this as the distinction between SF and AGN is solid in Table \ref{tab:table-AGN}. 
Another hint on the feasibility of our approach can be found in \citet{Khederlarian2024}, who used a neural network to predict the equivalent width of bright optical emission lines from the continuum, having the creation of mock catalogs in mind. That such approaches proved to be a success (also earlier work from \citealt{Beck2016} with PCA) is not a surprise if one considers how emission lines are produced in the first place. Nebular line emission is created when recombination occurs or when collisionally excited states in atoms or ions decay. This, on the other hand, depends on both the ionizing radiation field (ionization parameter, properties of the stars ionizing the gas) and the metallicity of the gas (metals act as coolants in a nebula) \citep{Kewley2019}. Photoionization codes in combination with stellar population synthesis like in \cite{Byler2017} were therefore capable of showing how emission line intensities correlate with the stellar population properties (age, metallicity). The latter also heavily affects the continuum shape and the broad-band photometry. 

\subsection{Disentangling color and spatial information}

Our generative AI learns the distribution of spectra conditioned on 5-band images with a resolution of 64x64, containing spatially resolved information (morphology) and pure color/magnitude information. The question arises: which of both is more important in successfully generating synthetic spectra?
To explore this, we train multiple CDMs with the same overall architecture as before but different input images by degrading their resolution and color information. Instead of the 64x64 images, stepwise coarser ones with 32x32, 16x16, 8x8, 4x4, 2x2 and finally 1x1 pixels are used. The overall area covered by the image (the FoV) remains the same, as well as the assignment of objects in the training, validation, and test set. Additionally, cases were considered where not all five bands were used; in the most extreme case, only the g-band alone was given to the generative AI. After training for each case, the spectra of the test set are predicted. 

\begin{figure*}[ht]
    \centering
    \includegraphics[width=\linewidth]{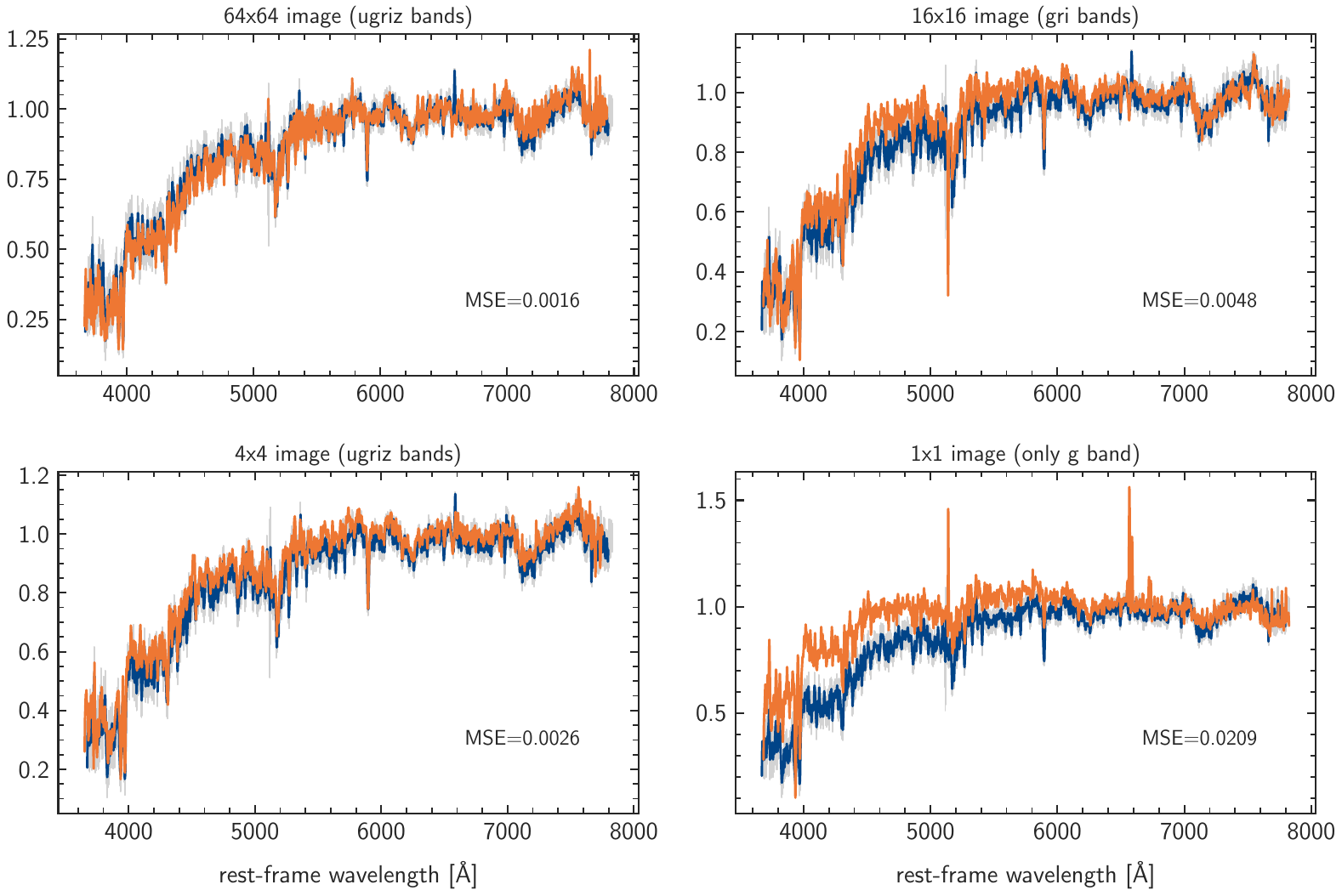}
    \caption{Generated spectra (in orange) for the object SDSS J091940.30+144450.4 based on input images with a resolution of 64x64 (in 5 bands), 16x16 (in 3 bands), 4x4 (in 5 bands), 1x1 (in 1 band only). The smoothed observed spectrum is drawn in each subplot in dark blue, and the 1-sigma error bars in grey. All spectra are shifted to the rest frame, and the MSE is written in each sub-figure for a better overview. Results are visibly better with higher-resolution images and more color information available. Given only the information in the g-band causes the generative AI to spuriously predict overall wrong shapes (wrong SEDs) and often strong emission lines.}
    \label{fig:quali}
\end{figure*}

\begin{figure*}[ht]
    \centering
    \includegraphics[width=0.8\linewidth]{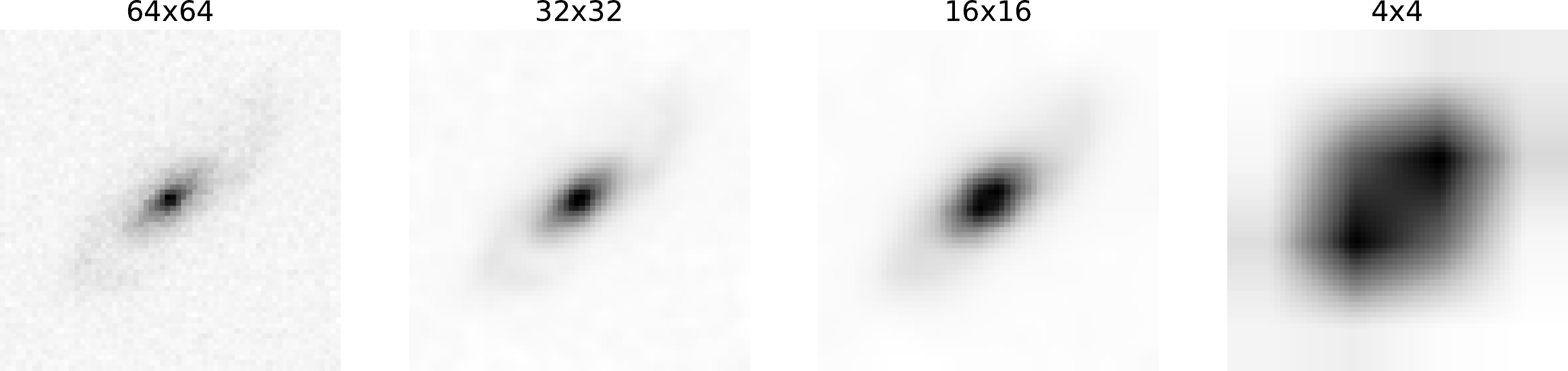}
    \caption{Illustration of the r-band of the spiral galaxy SDSS J143419.97-011332.1 using varying resolutions. This galaxy is part of the training set. } \label{fig:resolutions}
\end{figure*}

Not surprisingly, we find that by reducing the image resolution and thus gradually destroying the spatial information but keeping five bands, the mean squared error (MSE) of the generated spectra increases as listed in the first part of table \ref{tab:morphology}. This means that predicted spectra now differ more strongly from the observed ones. Also, having more bands available improves the performance of our method (second part of the table). Figure \ref{fig:quali} illustrates the differences in the generated spectra, showing results for different spatial resolutions and a different number of bands. We show an example of what the images look like after reducing resolution in Figure \ref{fig:resolutions}. 

Our experiment shows that both spatial information and magnitudes are important for the successful generation of spectra, although which is more important is inconclusive. Removing the two outermost bands has approximately the same effect as quartering the number of pixels in an image (one step coarser spatial resolution). Incorporating additional bands, especially small-band filters, will likely improve the results even further. Additional higher spatial resolution might not have such an impact.

Finally, using only one band is not sufficient enough to provide a usable spectrum, as the SEDs and emission lines strongly deviate from the real observations.

\begin{table}[htb]
\centering
\begin{tabular}{cc|c}
Resolution & Bands & MSE ($\times 10^{-3}$) \\
\hline
64x64 & 5 (ugriz) &  10.7 \\
32x32 & 5 (ugriz) &  10.8\\
16x16 & 5 (ugriz) &  11.5\\
8x8 & 5 (ugriz) &  12.4\\
4x4 & 5 (ugriz) &  14.3\\
2x2 & 5 (ugriz) &  20.0\\
1x1 & 5 (ugriz) &  24.7\\
\hline
32x32 & 3 (gri) &   11.6\\
16x16 & 3 (gri) &   12.2\\
\hline
1x1 & 1 (mean of ugriz) & 65.3 \\
1x1 & 1 (g) &  54.0\\
\end{tabular}
\caption{Mean MSE over 256 test objects with various image resolutions and bands (lower is better). All spectra are shifted to the rest-frame beforehand, and the fixed region of $4000-7000 \angstrom$ was used for this analysis. The numbers show that high-resolution, multi-band images give the best results. 
\label{tab:morphology}}
\end{table}

\subsection{Impact of the number of galaxies in the images}

\begin{figure*}[ht]
    \centering
    \includegraphics[width=\linewidth]{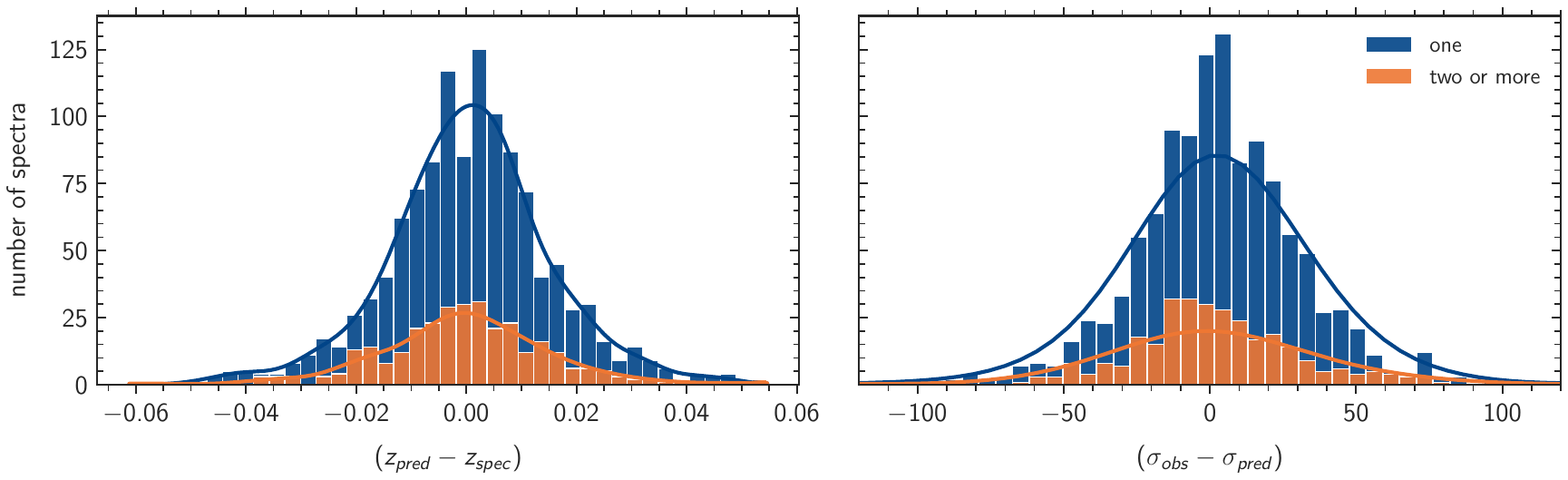}
    \caption{Distribution of the differences between the physical quantity derived from the observed and the predicted spectrum. Left: redshift prediction, right: velocity dispersion. The dark blue color means that only one galaxy is visible in the image, and the orange distribution means that two or more galaxies are clearly visible.}
    \label{fig:num_obs}
\end{figure*}

The training set is built in such a way that the spectroscopic target galaxy is always in the center of each image. However, in a substantial fraction ($\sim$22\%) of images, additional galaxies are visible near the borders. Can such a contamination lead to problems in redshift estimates or other physical properties? Or is it useful information? \\
To inspect this further, a reliable way of counting galaxies in the broad-band images is needed. We achieve this by using the tricolor g,r,i jpeg-images of each SDSS galaxy available on the SDSS website and working on them with the python \texttt{opencv} package. The process begins with converting each image to grayscale and applying subtle Gaussian blurs. Next, we utilize \texttt{opencv}'s built-in threshold function to transform the original image into a binary or segmented format using a predetermined threshold value. This step effectively distinguishes the galaxies (foreground) from the dark background. The final stage involves drawing and tallying contours, which provides the total galaxy count in each image. By adjusting and testing parameters at each processing step, we were able to develop a reliable method for determining the number of additional galaxies in various scenarios. To validate this approach, we manually counted several hundred galaxies as a cross-check. \\
It is after this step that it is possible to check whether the redshift information we obtain from the predicted spectra gets worse for multiple objects in the image. In figure \ref{fig:num_obs} (left part), we separated the test set into two different groups, depending on the number of overall visible objects in the images: only one galaxy or additional 'contamination' galaxies. The distributions for the various cases appear quite similar and lead to the same $\sigma_{MAD}$ values as calculated in the sections before. \\ 
A further research question might be whether other quantities might be affected by the contamination of additional galaxies. We checked all of them and only found no significant offset in any of the discussed quantities. Shown in figure \ref{fig:num_obs} (right part) is the velocity dispersion where one might expect a small effect given the distribution. In the median, the discrepancy is less than 3 km/s, negligible compared to the error estimate (around 8 km/s) from the bootstrapping procedure. Where it was possible, we also tested our results in the larger test set discussed in the appendix and came to the same conclusions.

\section{Summary} \label{sec:summ}

In this work, we have explored the possibility of generating complete galaxy spectra from photometric broadband images alone. We applied a versatile spectroscopic toolkit to evaluate the quality and information content of our ``artificial'' or ``predicted'' spectra, leading us to the following conclusions:

\begin{itemize}
    \item The mean relative deviation between the observed and predicted galaxy spectra showed the largest discrepancies in low-mass star bursting spirals due to wrong predictions of the extensive emission lines. A refined analysis incorporating the uncertainties of the observed flux at these data points showed that the galaxy spectra from different morphological groups were predicted equally well. 
    \item A comparison between spectral indices measured in predicted and observed spectra showed excellent agreement in Dn4000 and good agreement in the prominent spectral features [MgFe]$^\prime$, Hbeta, and Mg b. Difficulties for the generative AI arise only in weaker Lick indices such as Fe5270 and Fe5335. 
    \item The predicted and observed spectra were evaluated with two stellar population fitting codes (\texttt{pPXF} \& \texttt{FIREFLY}). The mean metallicity of the galaxy was recovered from the artificial spectrum; 86\% of all predicted spectra coincide in metallicity within 0.10 dex with \texttt{pPXF}. \texttt{FIREFLY} performed worse with 70\% agreement within 0.2 dex.  Overall, the use of \texttt{Firefly} delivered metallicities approximately 0.1 dex higher than those of \texttt{pPXF}. This is consistent with earlier findings from the literature comparing the performance of these two fitting codes. Values of the mean age of the stellar population and extinction showed good agreement: 0.3 dex scatter in log age and 0.1 dex scatter in E(B-V) as an overall rule of thumb.
    \item With our procedure generating artificial spectra, it becomes possible to predict the central velocity dispersions from photometric images alone. To our knowledge, this is the first attempt in the literature to do so. On our test set, we obtain values that are consistent with values from the observed spectra within 20\%.
    \item The presented machine learning algorithm recovers the famous bimodality of galaxy populations in colors, Dn4000, mean stellar age, extinction, and velocity dispersion using solely artificial spectra. It links colors and their 2D distribution (including morphological features) in the photometric images with physical quantities retrievable from spectra. 
    \item It is possible to identify AGN candidates from the photometric images with an accuracy of 82\%. For this, emission line ratios of the artificial spectra were evaluated in the BPT diagram. 
    \end{itemize}

In short, our approach can successfully predict various galaxy properties without explicitly training to do so. We show reasonable estimates for quantities such as age, metallicity, dust reddening, and velocity dispersion from photometric images alone by making a detour over artificial spectra, and more properties can easily be derived. The use of different fitting codes, different spectral templates, or the calculation of other physical quantities not mentioned here can be realized with the predicted spectra at free will - even years after the original generation of the spectra with the GenAI. This freedom in research justifies the computational costs. We believe the most promising applications of our method are in upcoming all-sky surveys such as Euclid and LSST, which will only have spectroscopic information on a small subset of the objects for which photometric images will be taken. By generating artificial spectra, we can, for instance, determine objects that are likely to be interesting and perform more targeted spectroscopic follow-ups. 

\section{Acknowledgments}
Special thanks go to Rolf-Peter Kudritzki and Luca Tortorelli for useful comments on the course of this work and the manuscript draft.
This work was funded by the Swiss National Science Foundation (SNSF), research grant 200021\_192285 “Image data validation for AI systems”. ES acknowledges support from the Computational Center for Particle and Astrophysics (C2PAP) as part of the Munich Excellence Cluster Origins funded by the Deutsche Forschungsgemeinschaft (DFG, German Research Foundation) under Germany’s Excellence Strategy EXC-2094 390783311. 
SC and MB acknowledge the ASI-INAF TI agreement, 2018-23-HH.0 ``Attività scientifica per la missione Euclid - fase D".
Funding for SDSS-III has been provided by the Alfred P. Sloan Foundation, the Participating Institutions, the National Science Foundation, and the U.S. Department of Energy Office of Science. The SDSS-III web site is \url{http://www.sdss3.org/}.
SDSS-III is managed by the Astrophysical Research Consortium for the Participating Institutions of the SDSS-III Collaboration including the University of Arizona, the Brazilian Participation Group, Brookhaven National Laboratory, Carnegie Mellon University, University of Florida, the French Participation Group, the German Participation Group, Harvard University, the Instituto de Astrofisica de Canarias, the Michigan State/Notre Dame/JINA Participation Group, Johns Hopkins University, Lawrence Berkeley National Laboratory, Max Planck Institute for Astrophysics, Max Planck Institute for Extraterrestrial Physics, New Mexico State University, New York University, Ohio State University, Pennsylvania State University, University of Portsmouth, Princeton University, the Spanish Participation Group, University of Tokyo, University of Utah, Vanderbilt University, University of Virginia, University of Washington, and Yale University.

\vspace{7mm}
\facilities{HST(STIS), Swift(XRT and UVOT), AAVSO, CTIO:1.3m,
CTIO:1.5m,CXO}


\software{Astropy \citep{astropy2013, astropy2018,astropy2022astropy},  
          Specutils affiliated with Astropy,
          FIREFLY \citep{Wilkinson2015, Wilkinson2017},
          pPXF \citep{Cappellari2004, Cappellari2017}
          PYPHOT \citep{Fouesneau2022},
          NumPy \citep{Numpy2020},
          Matplotlib \citep{matplotlib2007},
          SciPy \citep{scipy2021},
          OpenCV \citep{openCV1, openCV2}
          }


\bibliography{main}{}
\bibliographystyle{aasjournal}

\appendix \label{sec:appendix}
\vspace{-0.5cm}
\section{On the size of the test set} \label{app:size}
The size of a test set in machine learning is a crucial factor that can significantly influence model evaluation and performance assessment. In this paper, we worked with a relatively small test set of 1506 objects (0.5\% split). 
The bottlenecks of the whole analysis of the test set are the full-spectral fitting codes, which take of the order of minutes for one spectrum (including bootstrapping procedure). Using machine-learning techniques for the spectral evaluation seems promising, as recently demonstrated in \cite{Hunt2024}.
However, analyzing the performance of a black box ML algorithm with another new ML algorithm is too extreme and cannot be managed in one step. We believe it makes more sense to evaluate the predicted spectra in a traditional and well-founded manner. \\
To keep the computation times manageable, the presented test set was chosen quite small, but this always carries the danger of not having entirely representative performance estimates. In the following, we show that our results are nevertheless reliable by additionally evaluating the GenAI results on a test sample consisting of 14000 objects ($\sim$ 5\%) solely with \texttt{ppxf}. We chose the exact same settings as in the main part of the paper and recreated the figures \ref{fig:delta_cat}, \ref{fig:zpredict}, and \ref{fig:main_hist} now with this larger test set. Concerning the redshift estimates, we obtained a new value of $\sigma_{MAD}=0.01178$ compared to $\sigma_{MAD}=0.01177$ from the small test set. The obtained distributions look nearly identical compared to the ones of the smaller test set. We, therefore, conclude that it is unlikely that the smaller statistic hurts the results in the main body of the paper.

\begin{figure*}[htb]
        \medskip
	\center  \includegraphics[width=1\textwidth]{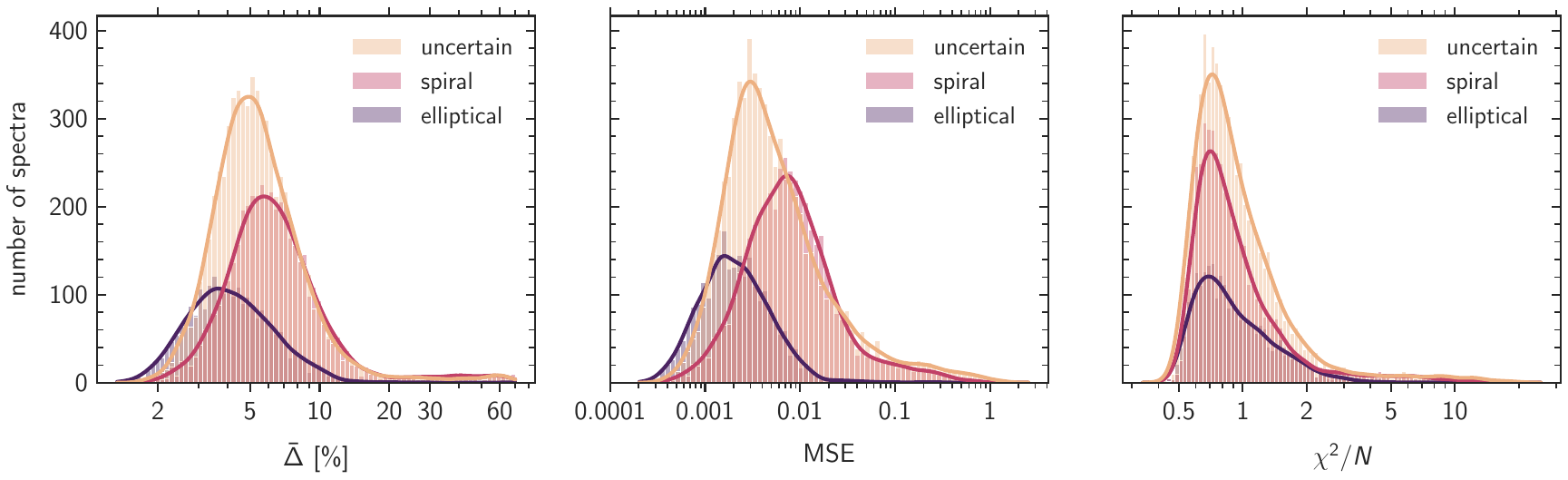}\medskip 
         \center \includegraphics[width=\textwidth]{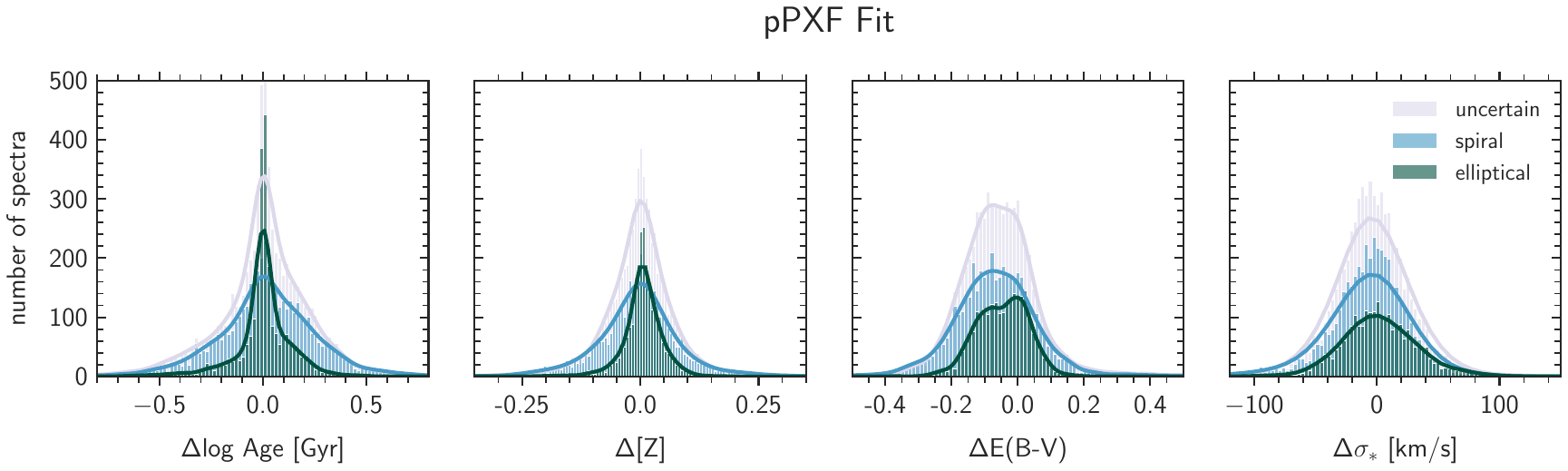}\medskip 
	\caption{Updated version of figure \ref{fig:delta_cat} and \ref{fig:main_hist} with the larger test set. Top: Histograms of $\overline{\Delta}$, Mean Square Error (MSE), and $\chi^2/N$ (with N the number of flux-points) of the 14000 galaxies split into the different morphologies. Bottom: Distribution of differences in the derived physical quantities for pPXF between the real and predicted spectra.}   
\end{figure*}

\begin{figure}[htb]
\vspace{0.3cm}
\center \includegraphics[width=0.4\linewidth]{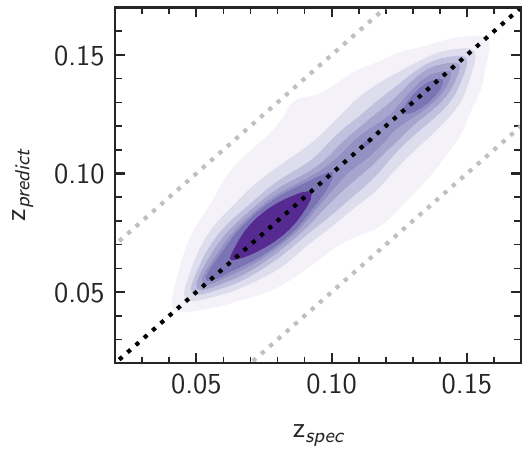}\medskip 
	\caption{Updated version of of figure \ref{fig:zpredict} with the larger test set. It shows the comparison between the spectroscopic redshift from the observed spectra (z$_{\textrm{spec}}$) and the prediction from our generative AI for the galaxies (z$_{\textrm{predict}}$). Instead of single data points, the overall kernel density estimate is shown for a clearer point of view. The grey dotted lines mark the regions with catastrophic outliers with $|\Delta z /(1+z_{spec})| >0.05$. The black dotted line shows the one-to-one relation.} 
\end{figure}

\section{Technical details} \label{app:tech}

In this section, we provide a more detailed description of our method, including technical aspects of diffusion models and contrastive learning, following the workflow in Figure \ref{fig:method}.

In the first step, we learn the conditional distribution over low-resolution spectra $p(\text{low-res spectrum}\mid\text{image})$ with a conditional diffusion model. \textbf{Diffusion models} are generative models that model the probability distribution of observed data and can generate new samples that closely mirror the characteristics of the training set. Diffusion models are trained using a combination of two Markov chains of length $T$, known as the \textit{forward} and \textit{backward} processes~\citep{Ho2020}. 

The forward process gradually adds Gaussian noise to a sample~$x_0$ with a fixed variance schedule $\beta_1,...,\beta_T$ depending on the time step $t \in \left[ 1,T\right]$. This process transforms the original sample $x_0$ into Gaussian noise at $x_T$. 
The backward denoising process aims to reverse the forward process by learning to predict the denoised version $x_0$ from $x_t$ with an autoencoder neural network $\epsilon_\theta$. In practice, $\epsilon_\theta$ learns to predict the noise \citep{Ho2020,Rombach2022}, with the simplified objective
\begin{equation}
    L_{DM} = \EX_{x_0,t,\epsilon\sim\mathcal{N}(0, \mathbf{I})} \left(\parallel \epsilon - \epsilon_\theta(x_t, t)\parallel^2 \right).
\end{equation}
After training is complete, the backward process describes the data distribution. With it, diffusion models can generate new samples by sampling from an isotropic Gaussian distribution and using the backward process to iteratively denoise it for $T$ timesteps. 

Vanilla diffusion models generate \textit{unconditional} samples. In our case, these will be spectra that are similar to the spectra in the training dataset, but they are not related to a specific image. Instead, in order to generate spectra for a specific object, we condition the noise prediction network $\epsilon_\theta$ on the input image. This allows us to sample from a \textit{conditional} distribution of spectra given an image. These models are known as conditional diffusion models (CDM). Specifically, the CDM projects the condition image $y$ to a latent space using a learnable mapping $\tau_\theta$ and introduces it into the network with cross-attention at multiple layers of the autoencoder~\citep{Rombach2022}. This gives the conditional loss function for a sample $x$,
\begin{equation}
    L_{CDM} = \EX_{x_0,y,t,\epsilon\sim\mathcal{N}(0, \mathbf{I})} \left(\parallel \epsilon - \epsilon_\theta(x_t, t, \tau_\theta(y))\parallel^2 \right),
\end{equation}
i.e., it learns to denoise a spectrum given the corresponding image.

Thus, the first step of our method is sampling multiple possible spectra for an object from the low-resolution CDM, CDM\textsubscript{lr}, given its image. However, as described in the main text, CDMs do not allow for density estimation and sampling from the CDMs results in multiple possible spectra for a given object without information on their likelihood. To select the most promising spectra for follow-up evaluation, we use multimodal contrastive learning~\citep{Chen2020} as a heuristic to find high-likelihood samples of the learned distribution.

\textbf{Contrastive learning} is a method for self-supervised learning. Unlike traditional supervised learning, which relies on manual annotations, in self-supervised learning, the supervision is automatically generated from unlabelled input data. For contrastive learning, the underlying idea is that two \textit{views} of a sample, such as two pictures of an object taken at different angles, should have a similar representation.

Formally, contrastive learning optimizes a neural network to minimize the distance between the features of two views of the same object while maximizing the distances to the features of other samples. This contrastive loss for a batch of size~$N$ is given by \citep{Chen2020}
\begin{equation}
\ell_{i, j} = -\log \frac{\exp (\text{sim}(\bm{z}_i, \bm{z}_j) / \tau)}{\sum^{2N}_{k=1} \bm{1}_{[k\neq i]} \exp (\text{sim}(\bm{z}_i, \bm{z}_k) / \tau)},
\end{equation}
for views $i$ and~$j$ of an object, where $\bm{z}$~represents their feature representation, $\tau$~the temperature, and $\text{sim}(\cdot,\cdot)$ a similarity measure, typically the cosine similarity.

In our case, the two views of an object are given by its image and spectrum. When the different views come from different modalities, this technique is called multimodal contrastive learning. In our case, we learn to map images and spectra into a shared representation space, where images and spectra with similar representations are likely to belong to the same object. 

The second step of our algorithm involves using the contrastive network trained on the low-resolution spectra and corresponding images to rank the generated spectra by CDM\textsubscript{lr} based on the similarities between their representations and that of the original image. We then select the best-matching samples to continue with the next steps. At this point, we have generated a handful of high-likelihood but low-resolution spectra for an object.

For step three, we train a second CDM to generate full-resolution spectra using the same process as in step one, with the only difference being an extra condition on the corresponding low-resolution spectrum. 
We do this by stacking the low-resolution spectrum with the original one channel-wise into $x_t^{comb}$, which is then used to train the full-resolution CDM:
\begin{equation}
    L_{CDM_{sr}} = \EX_{x_0,y,t,\epsilon\sim\mathcal{N}(0, \mathbf{I})} \left(\parallel \epsilon - \epsilon_\theta(x_t^{comb}, t, \tau_\theta(y))\parallel^2 \right).
\end{equation}
In short, this model learns to denoise a full-resolution spectrum, given the corresponding low-resolution spectrum and image, and is used to generate full-resolution versions of the most artificial spectra from step two.

The fourth step uses a second contrastive network to find the best matching full-resolution spectrum for an image. This contrastive network is trained as in step two, with the only difference being that it uses full-resolution instead of low-resolution images.

In summary, our full method samples several 563-dimensional spectra for an image with CDM\textsubscript{lr}. We select the three best synthetic spectra according to the low-resolution contrastive network. Then, we generate five full-resolution spectra for each of the selected low-resolution spectra. Finally, we select the best-matching spectrum with the full-resolution contrastive network, giving us the final synthetic spectrum for the object. 

\section{SQL Query} \label{app:sql}

{\scriptsize
\begin{verbatim}
    SELECT 
--/ Ids
 S.specObjID, SP.objID, P.ra, P.dec, 
  --/ Needed for spectra retrieving
 S.firstrelease, S.programname, S.instrument, S.run2d, 
 --/ Redshift and velocity dispersion
 S.z,  S.zerr, S.velDisp, S.velDispErr,
 --/ Petrosian radius
 P.petroRad_u, P.petroRad_g, P.petroRad_r, P.petroRad_i, P.petroRad_z, 
  --/ Radius containing 90% of Petrosian flux
 P.petroR90_u, P.petroR90_g, P.petroR90_r, P.petroR90_i, P.petroR90_z, 
 --/ Magnitude in 3 arcsec diameter fiber radius
 P.fiberMag_u, P.fiberMag_g, P.fiberMag_r, P.fiberMag_i, P.fiberMag_z, 
 --/ r-band extinction
 P.extinction_r, 
--/ Overall signal-to-noise-squared measure for plate 
 S.plateSN2, 
--/ surface brightness fit (exponential and de Vaucouleurs)
 P.modelMag_u, P.modelMag_g, P.modelMag_r, P.modelMag_i, P.modelMag_z, 
 P.cmodelMag_u, P.cmodelMag_g, P.cmodelMag_r, P.cmodelMag_i, P.cmodelMag_z, 
 --/ from galaxy zoo
 z.p_mg, z.p_el_debiased, z.p_cs_debiased, z.spiral, z.elliptical, z.uncertain,
 --/ photometric mass estimates
 M.logMass as logMass_Maraston09, CD.logMass as logMass_ConroyDust, CND.logMass as logMass_ConroyNoDust
 
INTO
 mydb.spectraGeneration
 
FROM 
 MyDB.MyTable_2 as SP 
 JOIN SpecObj as S ON SP.specObjID=S.specObjID
 JOIN PhotoObjAll as P ON SP.objID=P.objID
 JOIN zooSpec as Z ON SP.specObjID=Z.specObjID
 JOIN stellarMassFSPSGranWideDust as CD  ON SP.specObjID=CD.specObjID
 JOIN stellarMassFSPSGranWideNoDust as CND ON SP.specObjID=CND.specObjID
 JOIN stellarMassPassivePort as M  ON SP.specObjID=M.specObjID
WHERE 
 
 S.class = 'GALAXY' --/ select only galaxies (spectroscopically)
 AND P.mode = 1 --/ includes only objects which are “primary” in the survey
 AND S.z >= 0.05 --/ minimum redshift
 AND S.z<=0.15 --/ maximum redshift
 AND S.instrument = 'SDSS' --/ spectra from only one instrument
 
 --/ Removing Objects with Deblending Problems
 
 AND (flags_r & 0x20) = 0 --/ not PEAKCENTER 
 AND (flags_r & 0x80000) = 0 --/ not NOTCHECKED
 AND ((flags_r & 0x400000000000) = 0 OR P.psfmagerr_r <= 0.2) 
 --/ high S/N or not DEBLEND_NOPEAK this is suggested by SDSS 
 
 --/ Removing Objects with Interpolation Problems
 
 AND (flags_r & 0x800000000000) = 0 --/ not PSF_FLUX_INTERP
 AND (flags_r & 0x10000000000) = 0 --/ not BAD_COUNTS_ERROR
 AND ((flags_r & 0x100000000000) = 0 OR (flags_r & 0x1000) = 0) 
 --/ not both INTERP_CENTER AND CR
 
 --/ Removing Suspicious Detections
    --/ For stars AND galaxies (with type = 6 or 3), the "clean" flag checks 
    --/ that the object has pixels detected in the first pass (BINNED1), 
    --/ that it isn't saturated (SATURATED), AND that it has a valid radial 
    --/ profile (NOPROFILE).
 
 AND (flags_r & 0x10000000) != 0 --/ BINNED1
 AND (flags_r & 0x40000) = 0 --/ not SATURATED
 AND (flags_r & 0x80) = 0 --/ not NOPROFILE
 
  --/ Remove Meaningless Values

 AND P.petroRad_u>0
 AND P.petroRad_g>0
 AND P.petroRad_r>0
 AND P.petroRad_i>0
 AND P.petroRad_z>0

 AND P.petroR90_u>0
 AND P.petroR90_g>0
 AND P.petroR90_r>0
 AND P.petroR90_i>0
 AND P.petroR90_z>0
 
 AND P.petroRad_u<20
 AND P.petroRad_g<20
 AND P.petroRad_r<20
 AND P.petroRad_i<20
 AND P.petroRad_z<20
 
 AND P.petroR90_u<20
 AND P.petroR90_g<20
 AND P.petroR90_r<20
 AND P.petroR90_i<20
 AND P.petroR90_z<20

 AND P.fiberMag_u>0
 AND P.fiberMag_g>0
 AND P.fiberMag_r>0
 AND P.fiberMag_i>0
 AND P.fiberMag_z>0
 
 AND P.modelMag_u>0
 AND P.modelMag_g>0
 AND P.modelMag_r>0
 AND P.modelMag_i>0
 AND P.modelMag_z>0
  
 AND P.cmodelMag_u>0
 AND P.cmodelMag_g>0
 AND P.cmodelMag_r>0
 AND P.cmodelMag_i>0
 AND P.cmodelMag_z>0   
\end{verbatim}
}

\end{document}